\documentclass{caosp305R}
\usepackage{graphicx}
\usepackage{natbib}

\articleNo{269}
\pubyear{2016}
\volume{46}
\volnumber{2}
\firstpage{95}
\received{December 10, 2016}

\begin{document}

\newcommand{\zav}[1]{\left(#1\right)}
\newcommand{\hzav}[1]{\left[#1\right]}
\newcommand{\szav}[1]{\left\{#1\right\}}
\newcommand{\oc}{\textit{O-C}}
\newcommand\cir{BS\,Cir}
\newcommand\vir{CU\,Vir}
\newcommand\uma{CQ\,UMa}
\newcommand\ori{V901\,Ori}
\newcommand\sig{$\sigma$\,Ori\,E}
\newcommand{\teff}{$T_{\mathrm{eff}}$}
\newcommand{\mik}{Mikul\'{a}\v{s}ek}
\newcommand{\ziga}{\v{Z}i\v z\v novsk\'y}
\newcommand{\miket}{Mikul\'{a}\v{s}ek et al.}
\newcommand{\The}{\mathit{\Theta}}
\newcommand{\bet}{$\beta$}
\newcommand{\vt}{\tilde{\vartheta_0}}

\hauthor{Z.\,Mikul\'{a}\v{s}ek}
\htitle{Monitoring of rotational period variations in mCP stars}

\title{Monitoring of rotational period variations in~magnetic chemically peculiar stars}

\author{Zden\v{e}k Mikul\'{a}\v{s}ek}

\institute
    {Department of Theoretical Physics and Astrophysics, Masaryk University, Brno, Kotl\'a\v{r}sk\'{a} 2, CZ-611\,37 Brno, Czech Republic}

\date{December 10, 2016}

\maketitle

\begin{abstract}
A majority part of magnetic chemically peculiar (mCP) stars of the upper main sequence exhibits strictly periodic light, magnetic, radio, and spectral variations that can be fully explained by the model of a rigidly rotating main-sequence star with persistent surface structures and stable global magnetic field frozen into the body of the star. Nevertheless, there is an inhomogeneous group consisting of a few mCP stars whose rotation periods vary on timescales of decades, while the shapes of their phase curves remain nearly unchanged. Alternations in the rotational period variations, proven in the case of some of them, offer new insight on this theoretically unpredicted phenomenon. We present a novel and generally applicable method of period analysis based on the simultaneous exploitation of all available observational data containing phase information. This phenomenological method can monitor gradual changes in the observed instantaneous period very efficiently and reliably. We present up to date results of the period monitoring of V901~Ori, CU~Vir, $\sigma$\,Ori\,E, and BS~Cir, known to be mCP stars changing their observed periods and discuss the physics of this unusual behaviour. To compare the period behavior of those stars, we treated their data with an orthogonal polynomial model, which was specifically developed for this purpose. We confirmed period variations in all stars and showed that they reflect real changes in the angular velocity of outer layers of the stars, fastened by their global magnetic fields. However, the nature of the observed rotational instabilities has remained elusive up to now. The discussed group of mCP stars is inhomogeneous to such extent that each of the stars may experience a different cause for its period variations.

\end{abstract}

\section{Introduction}

It has been known since the times of the first systematic spectral classification that about 10\% of the upper main sequence stars show peculiar spectra with enhanced or extinct absorption lines of some chemical elements that indicate their overabundance/underabundance in respect to solar chemical composition. Some of those chemically peculiar (CP) stars have also been identified as variable stars exhibiting periodic moderate light variability with amplitudes up to one-tenth of the magnitude accompanied by variations in the intensities and profiles of spectral lines of some elements (as a rule overabundant ones). Spectropolarimetry of the variable CP stars has subsequently shown that the majority of variable CP stars have a strong, stable, nearly dipolar magnetic field, whose axis does not coincide with the rotational one.

The magnetic chemically peculiar (mCP) stars are divided into several subtypes, namely classical Ap and Bp stars with an overabundance of the iron peak elements, silicon, strontium or rare earths, and hot mCPs with underabundance/overabundance of helium (He-weak and He-strong stars). The overabundant chemical elements in their atmospheres usually concentrate into large spots, stable for decades or centuries. The unequal distribution of chemical elements on the surface influences the horizontal inhomogeneity of their atmospheric structure resulting into the incidence of extended photometric spots with uneven energy distributions in their spectra \citep{krt901,shul,krtcu}. As the star rotates, periodic variations in the brightness, spectrum, and magnetic field are observed. The period of the observed variations is equal to rotational period. Combining both present and archival observations of mCP stars collected over the past several decades, we can reconstruct their rotational evolution with unprecedented accuracy.

\section{Models for period monitoring}

During the last two decades, the research team around the Department of Theoretical Physics and Astrophysics of the Masaryk University in Brno has developed several versatile instruments for the analysis of the periodicity of more or less periodically variable objects, using the methods of phenomenological modelling. In this section, we briefly outline the period analysis technique apt for an investigation of the period stability of magnetic chemically peculiar stars.

\subsection{Phase function and its models}\label{phasefunsec}

Most of the variations in periodical variable stars are cyclic with an instantaneous period $P(t)$, which is strictly constant or slightly variable with time $t$. The period itself and its progression over time cannot be observed directly, but both
can be derived through analysing time series of light changes or extremum timings. For that purpose we introduced \citep[see][]{mik901,mikecl}) a monotonically rising \emph{phase function} $\vartheta(t)$ as a sum of the epoch $E(t)$ and the common phase $\varphi(t)$ and its inversion function $t(\vartheta)$,
\begin{equation}
\vartheta=E+\varphi;\quad \varphi=\mathrm{FP}(\vartheta);\quad E=\mathrm{IP}(\vartheta),
\end{equation}
where $\mathrm{IP}(x)$ is an operator rounding $x$ to the nearest integer less than or equal to $x$, while $\mathrm{FP}(x)=x-\mathrm{IP}(x)$. The phase function equals zero at the time of the initial epoch, denoted usually by $M_0$, hence $\vartheta(t=M_0)=0$. Using the inversion function $t(\vartheta)$ we can find for any value of the phase function the corresponding time. The discrete form of this function $t(E)$ predicts the time of the zero phase appertained to the particular epoch $E$.

Functions $\vartheta(t)$ and $t(\vartheta)$ are tied with the instantaneous observed periods $P(t),\,P(\vartheta)$ by the following differential equations \citep[for details see][]{mikecl}
\begin{eqnarray}\label{phasefundef}
\displaystyle\frac{\mathrm{d}\vartheta(t)}{\mathrm{d}t}=\frac{1}{P(t)}\quad \vartheta(M_0)=0;\quad \vartheta=\int_{M_0}^t \frac{\mathrm d \tau}{P(\tau)}; \\ \displaystyle\frac{\mathrm{d}t(\vartheta)}{\mathrm{d}\vartheta}=P(\vartheta);\quad  t(\vartheta) =M_0+\int_0^\vartheta P(\zeta)\ \mathrm d \zeta, \nonumber
\end{eqnarray}

First of all, we remind that the optimal option of the model of the phase function $\vartheta(t)$ (and its inversion function $t(\vartheta)$) is crucial for the whole period analysis. The parameters of the model are iteratively determined by the process of modeling of the observed behavior of the variable star. The time dependence of the instantaneous period $P(t)$ or $P(\vartheta)$ is then a function derived from $\vartheta(t)$ according to general relations given by Eq.\,(\ref{phasefundef}). So, the function $P(t)$ strongly depends on the chosen model of the phase function and its adequacy.

The basic and at the same time the simplest model of the variability supposes that the observed period of variations is punctually constant $P(t)=P(\vartheta)=P(\The)=P_0$. Using Eq.\,(\ref{phasefundef}) we get the corresponding, linear phase function $\vartheta_0(t)$ and its inversion $t(\vartheta_0)$, described by two parameters $(M_0,P_0)$:
\begin{equation}\label{linaprox}
\displaystyle\vartheta_0(t)=\int_{M_0}^t \frac{\mathrm d \tau}{P_0}= \frac{t-M_0}{P_0};\quad
t(\vartheta_0) = M_0+P_0\,\vartheta_0;\quad \frac{\mathrm dt}{\mathrm d\vartheta_0}=P_0.
\end{equation}

Because $\vartheta_0$ is a linear function of time, we can use it instead of time. Combining Eq.\,(\ref{phasefundef}) and Eq.\,(\ref{linaprox}) we obtain the useful relations
\begin{equation}\label{combine}
\displaystyle P(\vartheta_0)=P_0\hzav{\frac{\mathrm{d}\vartheta(\vartheta_0)} {\mathrm{d}\vartheta_0}}^{-1};\quad P(\vartheta)=
P_0\,\frac{\mathrm{d}\vartheta_0(\vartheta)}{\mathrm{d}\vartheta}.
\end{equation}

\subsubsection{Power law models}

Several mechanisms, as a steady angular momentum loss via stellar winds or magnetic breaking, subsequently change the instantaneous periods of astrophysical objects $P(t)$, or their frequencies $\nu(t)=1/P(t)$, according to the so-called \emph{power law}, as follows
\begin{equation}\label{power}
\dot{\nu}(t)=-K\,\nu^q \quad \Rightarrow \quad \dot{P}(t)=K\,P^{2-q},
\end{equation}
where $q$ is the so-called \emph{deceleration parameter}, symptomatic for the dominating mechanism of the period change and $K$ is a proportionality constant characteristic of a particular object. If $K>0$, the frequency declines while the period is rising.
\begin{eqnarray}\label{qecko}
&\displaystyle \quad K=-\frac{\dot{\nu}}{\nu^q}=-\frac{\dot{\nu}_0}{\nu_0^q}=
\frac{\dot{P}}{P^{2-q}}=\frac{\dot{P}_0}{P^{2-q}_0}\\
&\displaystyle \dot{\nu}=\dot{\nu}_0\zav{\frac{\nu}{\nu_0}}^q, \quad \frac{\mathrm d\nu}{\mathrm d\vartheta_1}=\frac{\dot{\nu}_0}{\nu_0}\zav{\frac{\nu}{\nu_0}}^q, \quad \ddot{\nu} =q\zav{\frac{\dot{\nu}_0}{\nu^q_0}}^2\nu^{2\,q-1}, \quad \Rightarrow \quad \frac{\nu\,\ddot{\nu}}{\dot{\nu}^2} =q, \nonumber \\
&\displaystyle \dot{P}=\dot{P}_0\zav{\frac{P}{P_0}}^{2-q}, \  \frac{\mathrm dP}{\mathrm d\vartheta_1}=\dot{P}_0\,P_0\zav{\frac{P}{P_0}}^{2-q}, \  \\
&\displaystyle \ddot{P} =(2-q)\zav{\frac{\dot{P_0}}{P^{2-q}_0}}^2 P^{3-2\,q}, \quad  \frac{P\,\ddot{P}}{\dot{P}^2} =\frac{P_0\,\ddot{P}_0}{\dot{P}_0^2}=2-q, \nonumber
\end{eqnarray}
where $\dot{P}_0\ \mathrm{and}\ \ddot{P}_0$ are the first at $t=M_0$.

Because the secular changes are typically slow, we can approximate the phase function $\vartheta(\vartheta_0)$ and its inversion $\vartheta_0(\vartheta)$ using the Maclaurin expansion up to the cubic term. After some algebra, we obtain the following relations,
\begin{equation}
\vartheta(\vartheta_0)\doteq\vartheta_0-\frac{\dot{P}_0}{2}\,\vartheta_0^2+ \frac{q\,\dot{P}_0^2}{6}\,\vartheta_0^3,\quad \vartheta_0(\vartheta)\doteq\vartheta+\frac{\dot{P}_0}{2}\,\vartheta^2+ \frac{(3-q)\,\dot{P}_0^2}{6}\,\vartheta^3.
\end{equation}
The physically interesting parameter $q$ is presented only in cubic terms with the connection to the square of the period progression $\dot{P}_0^2$, which is almost always negligible\footnote{\ori\ shows, among other period changing mCPs, a record value of rotational braking $\dot{P}=1.0\times10^{-8}$. The cubic correction of the $\mathit{\The(E)}$ in the time of maxima even during 100 years does not exceed 1 minute, while the uncertainty of its determination from observations is always larger than 10 minutes.}. Consequently, we have to reconcile the impossibility to determine the value of the deceleration parameter from observations and to reveal the background of the mechanisms causing a constant rise of the period. However, there are objects with rotational variability displaying cubic and higher terms. Nevertheless, these variations in their periods have to be caused by mechanisms not obeying a power law.

\subsubsection{Standard Mclaurin polynomials}\label{maclaurinpol}

Let us admit that the observed phase curves display systematical phase (or time) shifts versus their linear model prediction amounting to the $-\Delta(\vartheta_0)$ or $P_0\,\Delta(\vartheta)$ difference. We can interpret this as a result of the inconstancy of the instantaneous period $P(\vartheta_0)$ or $P(\vartheta)$.
\begin{eqnarray}
&\displaystyle \vartheta(\vartheta_0)=\vartheta_0-\Delta(\vartheta_0);\quad
\vartheta_0=\vartheta+\Delta(\vartheta_0)\doteq\vartheta+\Delta(\vartheta)+ \frac{1}{2}\frac{\mathrm d\Delta^2}{\mathrm d\vartheta}+\ldots;\\
&\displaystyle P(\vartheta_0)=P_0\hzav{\frac{\mathrm{d}\vartheta(\vartheta_0)} {\mathrm{d}\vartheta_0}}^{-1}=\frac{P_0}{\displaystyle 1-\frac{\mathrm d\Delta}{\mathrm d\vartheta_0}}\doteq P_0\hzav{1+\frac{\mathrm d\Delta}{\mathrm d\vartheta_0}+\zav{\frac{\mathrm d\Delta}{\mathrm d\vartheta_0}}^2+\ldots};\\ & \displaystyle P(\vartheta)\doteq P_0\hzav{1+\frac{\mathrm d\Delta}{\mathrm d\vartheta}+\frac{1}{2}\frac{\mathrm d^2\Delta^2}{\mathrm d\vartheta^2}+\ldots}.
\end{eqnarray}
Then, we can simply write:
\begin{eqnarray}\label{simple}
&\displaystyle \vartheta(\vartheta_0)=\vartheta_0-\Delta(\vartheta_0);\
\vartheta_0(\vartheta)=\vartheta+\Delta(\vartheta);\ \The(E)=M_0+P_0\,\hzav{E+\Delta(E)};\\
&\displaystyle P(\vartheta_0)\doteq P_0\zav{1+\frac{\mathrm d\Delta}{\mathrm d\vartheta_0}}\doteq P(\vartheta)=P_0\zav{1+\frac{\mathrm d\Delta}{\mathrm d\vartheta}}=P(E)=P_0\zav{1+\frac{\delta\Delta}{\delta E}},\nonumber
\end{eqnarray}
where $\The(E)$ is a model prediction of the time of the zeroth phase $(\varphi=0)$ corresponding to the epoch $E$.

In particular, if we express $P(t)$ by means of the Mclaurin expansion where $\dot{P}_0\ \mathrm{and}\ \ddot{P}_0$ are the first and second time derivatives at the time of the origin $t=M_0$, which may be fixed. Then
\begin{eqnarray}\label{thetalaurin}
& \Delta(\vartheta_0)=\frac{1}{2!}\dot{P}_0\vartheta_0^2+ \frac{1}{3!}P_0\ddot{P}_0\vartheta_0^3+\ldots+\frac{1}{(k+1)!}\,P_0^{k-1} \frac{\mathrm d^k\!P_0}{\mathrm dt^k}\,\vartheta_0^{k+1}+\ldots; \\
&\vartheta(\vartheta_0)=\vartheta_0 -\Delta;\quad\vartheta_0(\vartheta) \doteq \vartheta+\Delta;\quad \The(E)\doteq M_0+P_0[E+\Delta(E)];\nonumber \\
&\displaystyle P(\vartheta_0)=P_0\zav{1+\dot {P}_0\,\vartheta_0+\textstyle{\frac{1}{2}}P_0\ddot{P}_0\,\vartheta_0^2+\ldots+ \frac{1}{k!}P_0^{k-1} \frac{\mathrm d^k\!P_0}{\mathrm dt^k}\,\vartheta_0^{k}+\ldots};\nonumber\\
&\dot{P}(\vartheta_0)=\frac{\mathrm dP}{\mathrm d\vartheta_0}\frac{\mathrm d\vartheta_0}{\mathrm d t} =\dot{P}_0+P_0\ddot{P}_0\,\vartheta_0+\ldots+\frac{1}{(k-1)!}\,P_0^{k-1} \frac{\mathrm d^k\!P_0}{\mathrm dt^k}\,\vartheta_0^{k-1}+\ldots\nonumber\\
&\ddot{P}(\vartheta_0)=\frac{\mathrm d\dot{P}}{\mathrm d\vartheta_0}\frac{\mathrm d\vartheta_0}{\mathrm d t} =\ddot{P}_0+\ldots+\frac{1}{(k-2)!}\,P_0^{k-2} \frac{\mathrm d^k\!P_0}{\mathrm dt^k}\,\vartheta_0^{k-2}+\ldots\nonumber
\end{eqnarray}

\subsubsection{Cyclic variation of the period}\label{cyclic}

The modulation amounting to $\Delta(\vartheta_0)$ of the ideal linear phase function $\vartheta_0(t)$ may represent a curving of it, expressing a subsequent change of the period, or may be cyclic\footnote{Note that the possible variations in the observed period of mCP star variations need not necessarily mean changes in the rotational period itself. They can also be caused by the inconstant radial velocity of the star as the result of orbital motion in a stellar system. Orbital motion results in some undulation of the basic phase function causing the light-time effect \citep[for details see][]{liska}.}, with a proper period $\mathit{\Pi}$, different from the basic, in our case rotational period $P(\vartheta_0)$.

Let us assume that the difference $\Delta(\phi)$ is a periodic function of the time dependent variable $\phi(t)=(t-T_0)/\mathit{\Pi}$, where $T_0$ is the origin of counting of cycles with the period $\mathit{\Pi}$. Then, we can proceed according to the following algebra
\begin{eqnarray}\label{cycpsi}
&\displaystyle t=M_0+P_0\,\vartheta_0;\quad \phi(t,\vartheta_0)= \frac{t-T_0}{\mathit{\Pi}}=\frac{P_0\,\vartheta_0+M_0-T_0}{\mathit{\Pi}}; \quad \Delta(\phi)=\Delta(\vartheta_0);\nonumber\\
&\vartheta=\vartheta_0-\Delta(\vartheta_0);\quad \The_0(E)=M_0+P_0 E;\quad \The(E)=\The_0(E)+P_0\Delta(E);\\
&\displaystyle P(\vartheta_0)=P_0\zav{1+\frac{\mathrm d\Delta}{\mathrm d\vartheta_0}}=P_0\zav{1+\frac{\mathrm d\Delta}{\mathrm d\phi}\frac{\mathrm d\phi}{\mathrm d\vartheta_0}}=P_0\zav{1+\frac{P_0}{\mathit {\Pi}}\frac{\mathrm d\Delta}{\mathrm d\phi}}. \nonumber
\end{eqnarray}
If we are allowed to opt for the time of the origin of the cyclic phase function $\phi$, $T_0$, and the definition of the cyclic function $\Delta(\phi)$, it is advantageous to choose them so that: $\Delta(\vartheta_=0)=0$ and $\dot{\Delta}(\vartheta_=0)=0$, then $\vartheta(t=M_0)=0$ and $P(t=M_0)=P_0$. Please also refer to Sect.\,\ref{cuvir}, where we apply the set of relations (\ref{cycpsi}) to the simple sinusoidal modulation of the basic phase function.

\subsubsection{Orthogonalized models}\label{ortopol}

The most widely used models of phase function $\vartheta$ are expressed as linear combinations of a basic set of functions of $\vartheta_0$: $\szav{\psi_0,\,\psi_1,\,\psi_2,\,\psi_3,\ldots}$. For several good reasons \citep[see e.\,g.][]{mikort,mik901}, it is advantageous to switch from that to another set of
functions $\szav{\theta_0,\,\theta_1,\,\theta_2,\,\theta_3,\ldots}$ which are mutually orthogonal in the space of measurements. This means that the weighted mean value of the products of each two uneven functions is equal to zero: $\overline{\theta_j\,\theta_k}=0,\ \mathrm{for}\ j\neq k$. The functions of this set can be created iteratively by means of the well-known Gram-Schmidt procedure as the linear combination of the former set $\szav{\psi_0,\,\psi_1,\,\psi_2,\,\psi_3,\ldots}$ as follows
\begin{eqnarray}\label{ortog}
&\displaystyle \theta_0=\psi_0;\quad \theta_1=\psi_1-\alpha_{10}\psi_0;\quad
\alpha_{10}=\frac{\overline{\psi_1\,\theta_0}} {\overline{\theta_0^2}}; \nonumber \\
&\displaystyle\theta_2=\psi_2-\alpha_{20}\theta_0-\alpha_{21}\theta_1;\quad \alpha_{20}=\frac{\overline{\psi_2\,\theta_0}} {\overline{\theta_0^2}}; \quad \alpha_{21}=\frac{\overline{\psi_2\,\theta_1}} {\overline{\theta_1^2}}; \nonumber \\
&\displaystyle\theta_3\!=\!\psi_3\!-\!\alpha_{30}\theta_0\!-\!\alpha_{31}\theta_1\!-\!
\alpha_{32}\theta_2;\ \alpha_{30}\!=\!\frac{\overline{\psi_3\theta_0}} {\overline{\theta_0^2}};\ \alpha_{31}\!=\!\frac{\overline{\psi_3\theta_1}} {\overline{\theta_1^2}};\ \alpha_{32}\!=\!\frac{\overline{\psi_3\theta_2}} {\overline{\theta_2^2}}; \nonumber \\
&\displaystyle \theta_j=\psi_j-\sum_{k=0}^{j-1}\alpha_{jk}\,\theta_k;\quad \alpha_{jk}=\frac{\overline{\psi_j\,\theta_k}}{\overline{\theta_k^2}}.
\end{eqnarray}
Although this way of expressing of newly created orthogonal functions is suitable for computation, it would be more illustrative to rewrite the relations for particular orthogonal terms $\theta_j$ using solely the functions $\psi_k$, where $k=(0,1,2,\ldots,j)$.
\begin{eqnarray}\label{ortobeta}
&\theta_0=\psi_0;\quad  \theta_1=\psi_1+\beta_{10}\psi_0,\quad \beta_{10}=\alpha_{10};\\
&\theta_2=\psi_2-\beta_{21}\psi_1-\beta_{20}\psi_0,\quad \beta_{21}=\alpha_{21},\quad\beta_{20}=\alpha_{20}-\alpha_{21}\alpha_{10}; \nonumber\\
&\theta_3=\psi_3-\beta_{32}\psi_2-\beta_{31}\psi_1-\beta_{30}\psi_0; \quad
\beta_{32}=\alpha_{32},\quad\beta_{31}=\alpha_{31}-\alpha_{32}\alpha_{21},\nonumber\\
&\beta_{30}=\alpha_{30}-\alpha_{31}\alpha_{10}-\alpha_{32}\alpha_{20}+ \alpha_{32}\alpha_{21}\alpha_{10};
\nonumber\\
&\theta_4=\psi_4-\beta_{43}\psi_3-\beta_{42}\psi_2-\beta_{41}\psi_1- \beta_{40}\psi_0;\ \beta_{43}=\alpha_{43},\ \beta_{42}=\alpha_{42}- \alpha_{43}\alpha_{32},\nonumber\\
&\beta_{41}=\alpha_{41}-\alpha_{42}\alpha_{21}-\alpha_{43}\alpha_{31}+ \alpha_{43}\alpha_{32}\alpha_{21},\nonumber\\
&\beta_{40}\!=\!\alpha_{40}\!-\!\alpha_{43}\alpha_{30}\!-\! \alpha_{42}\alpha_{20}\!-\!\alpha_{41}\alpha_{10}\!+
\!\alpha_{43}\alpha_{32}\alpha_{20}\!+\!\alpha_{42}\alpha_{21}\alpha_{10}-
\alpha_{43}\alpha_{32}\alpha_{21}\alpha_{10}\nonumber,
\end{eqnarray}
and so on. The general repetitive orthogonalization routine, as described by Eq.\,(\ref{ortog}), can be applied to the standard polynomial set of $\vartheta_0$, as it has been shown in Sect.\,\ref{ortogpolmodel}.

Similarly, we can also orthogonalize models of the phase function $\vartheta$ containing non-polynomial terms, including the cyclical ones (see Sect.\,\ref{cyclic}). The simplest possibility is a model with the constant basic period $P_0$ and some phase cosine-like modulation as a function of the phase $\phi$ (see Eq.\,(\ref{cycpsi})) with the set of elementary functions $\{1,\ \vartheta_0,\ \cos(2\pi\phi)\}$. Assuming that $\tilde{M}_0$ has been chosen so that $\overline{\vartheta_0}=0$, we arrive at the following model
\begin{equation}
\vartheta_0=\frac{t-\tilde{M}_0}{P_0};\quad\vartheta=\vartheta_0-A\hzav{\cos(2\pi\phi) -\frac{\overline{\vartheta_0\cos(2\pi\phi)}}{\overline{\vartheta_0^2}}\  \vartheta_0-\overline{\cos(2\pi\phi)}}.
\end{equation}

\subsection{Phenomenological models of mCP phase curves}\label{phasecurves}

The development of the period $P(t)$ of a mCP star can be derived from the mutual phase shifts among phase curves obtained at different times, where $\Delta \varphi=\varphi(t_2)-\varphi(t_1)=\vartheta(t_2)-\vartheta(t_1)- [\vartheta_0(t_2)-\vartheta_0(t_1)]$. For the determination of the phase shifts, we use the fact that the phase curves of mCP stars remain invariable for many decades. To obtain a maximum interval of observations we are forced to combine all available phase curve of varying characteristics of the star. It is advantageous to use proper phenomenological models of phase curves here, describing them accurately, using a minimum of free parameters.

The following review of the phase curve models is only rough and incomplete because the observed mCP star variations are in reality extremely rich and diverse. Therefore, it is common to tailor the models differently for each star.

\subsubsection{Light curves}\label{lightcurves}

The period analyses of mCP stars are based mainly on photometry. Monochromatic light curves of mCP stars are the result of the presence of dull photometric spots on the surface of rotating stars. The resulting light curves are smooth and relatively simple. A majority of monochromatic light curves can be well approximated by the harmonic polynomial of the second or third order \citep{mik07b,jagmik}. It is advantageous to centre the light curve into the phase of light maximum $\varphi_m$, which uses to be sharper than the minimum/minima. The model of a monochromatic light curve then has maximally five phase dependent components -- three symmetric and two antisymmetric ones:
\begin{eqnarray}\label{mono}
&\displaystyle m(\vartheta)= m_0+ A_1 \cos(2\pi\varphi_m)+ A_2 \cos(4\pi\varphi_m)+ A_3
\cos(6\pi\varphi_m)+\\
&A_4\hzav{2 \sin(2 \pi \varphi_m)\!-\!\sin(4 \pi \varphi_m)}
+A_5\hzav{3 \sin(2\pi\varphi_m)\!+\!6\sin(4\pi\varphi_m)\!-\! 5\sin(6\pi\varphi_m)},\nonumber\\
&\mathrm{where}\quad\varphi_m=(\vartheta-\varphi_{0m})- \mathrm{round}(\vartheta-\varphi_{0m}).\nonumber
\end{eqnarray}
However, the shapes of the spectral energy distribution of individual photometric spots are often dissimilar, which results in the fact that light curves in various colours are not the same; they differ both in their amplitudes and forms. Fortunately, the usage of the PCA or APCA technique \citep{mik04,apca} considerably diminishes the number of parameters indispensable for good description of light curves obtained in all colours.

There is also another, more physical approach to the modelling of mCP light curves in various colors. It assumes that light curves in different colors can be expressed as a linear combination of a few symmetric basic profiles with that centers at phases $\vartheta_{0j}$ and half-widths $d_j$:
\begin{eqnarray}\label{cosh}
&\displaystyle m(\lambda,\varphi)= m_0(\lambda)+\sum_{j=1}^{n_{\mathrm{s}}}A_j(\lambda) \,\szav{\exp\hzav{1\!-\!\cosh\zav{\frac{\Delta \varphi_j}{d_j}}}-2.289\,d_j};\\
&\mathrm{where}\quad\Delta\varphi_j=(\vartheta-\varphi_{0j})- \mathrm{round}(\vartheta-\varphi_{0j}).\nonumber
\end{eqnarray}
The above mentioned phenomenological model applies to almost all mCP stars observed from ground-based observatories \citep[see e.\,g.][]{mik15,krtheta,jagmik}. Nevertheless, \sig\  forms an exception. Its light curves in the optical region can not be explained by the model of a rotating star with photometric spots \citep[][and references therein]{oxaiaus,oxa15}.

Most of the photometric data on the here studied mCP stars are freely accessible through the database {\it On line catalogue of photometric observations of magnetic chemically peculiar stars}, http://astro.physics.muni.cz/mcpod/ \citep{mikan,jan}.

\subsubsection{Magnetic fields}\label{magcurves}
Magnetic fields of mCP stars are structurally much simpler, and mainly much stronger, than the fields of cool stars \citep{dolan}. The large-scale strength and geometry of the magnetic fields are stable, in the rotating stellar reference frame, on time scales of many decades \citep[e.\,g.][]{wade00,silvester}. Magnetic field observations are then very reliable subsidiary source of the phase information on the rotation of mCP stars. A major part of practically exploited data is time-series of the observations of the mean longitudinal component $B_{\rm{eff}}$ of the total magnetic field. They were derived from measurements of circular polarization induced in magnetically-split spectral line $\sigma$ components due to the longitudinal Zeeman effect \citep[see, e.\,g.][]{mathys,dolan}.
\begin{figure}[h]
\centerline{\includegraphics[width=0.95\textwidth,clip=]{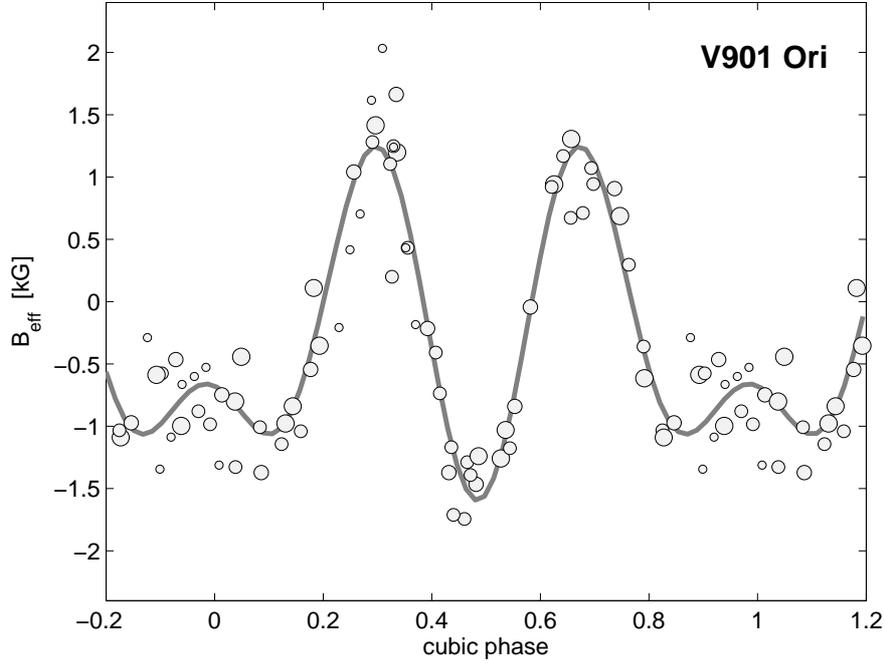}}
\begin{center}
\caption{\small The extraordinary phase curve of the mean longitudinal magnetic field of \ori, displaying three waves. The curve was plotted on the cubic phase (more information in Sect.\,\ref{901ori}). It was fitted with the symmetrical harmonic polynomial of the third order. However, single wave curves are typically observed in the vast majority of mCP stars.} \label{901beff}
\end{center}
\end{figure}

The magnetic fields of most of the mCP stars are more or less dipole-like with the dipole axis tilted to the rotational axis. The phase curves of the mean longitudinal component of the field $B_{\rm{eff}}$ are almost sinusoidal. However, the current spectropolarimetric measurements of high quality show that a quadruple component of the field of many mCP stars has to be considered \citep{wade00}. Nevertheless, a majority of magnetic phase curves are single waves, which can be described by a harmonic polynomial of maximally second order. An example of an mCP star harboring an extraordinarily complex magnetic field is one of our target stars -- \ori\ \citep{thomla,ko901}, the phase curve of which is only adequately described by the symmetric harmonic polynomial of the third order (see Fig.\,\ref{901beff}).

\subsubsection{Spectral line variations}\label{spectralcurves}

One of the specific features of mCP stars is a very uneven horizontal distribution of chemical elements on the stellar surface, which applies to those elements that are overabundant in respect to the solar abundance. Chemical elements concentrate to vast spots whose projection on the visible disc of the star changes as the star rotates. Owing to this, the spectral absorption lines strongly variable in their intensity and profiles with the rotational period of the star. Changes in equivalent widths of spectral lines of mCP stars are so conspicuous that they were noticed and measured at the very beginning of the study of mCP stars \citep[see e.\,g.][]{belopol,farn,deutsch}.

\begin{figure}[h]
\centerline{\includegraphics[width=0.99\textwidth,clip=]{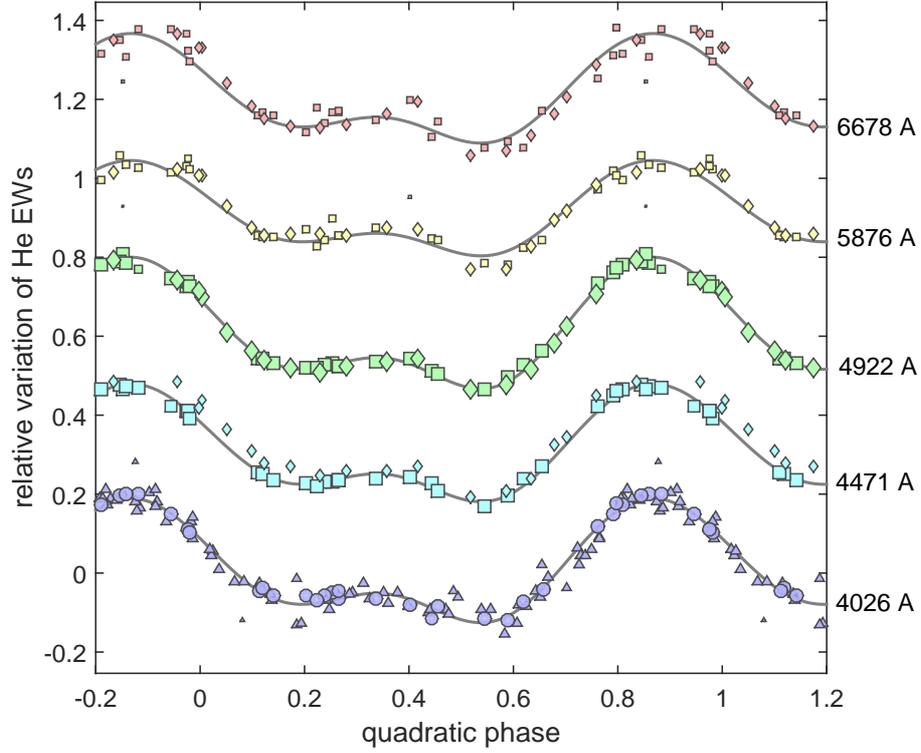}}
\begin{center}
\caption{\small The dependence of relative variations of equivalent widths of selected He\,I spectral lines of \sig\ versus the quadratic phase (see Sect.\,\ref{901ori}). The different symbols distinguish different sources of data, the areas of symbols are proportional to the weights of individual measurements. The solid lines denote the fits by the phenomenological model given by Eq.\,(\ref{spmodel}).} \label{sighe}
\end{center}
\end{figure}
The analysis of phase variation of the line profiles of ions of chemical elements, combined with changes in their polarization by the techniques of magnetic Doppler imaging, enables a successful mapping of the distribution of spectral spots in the atmospheres \citep[see e.\,g.][]{rice,choch,luft}. The pioneer systematic studies of the variability of He I and Si II lines in upper main sequence stars are those of \citet{ped}, which helped us very much in anchoring our period research in the eighties of the last century. Of a rather limited use are data on radial velocity variations because they depend strongly on the measured spectral line selection and the method of measurements. Relatively reliable are the radial velocity of the hydrogen Balmer lines or other strictly defined spectral features. Nevertheless, the basic phase information hides present and archival equivalent width measurements of selected lines of overabundant chemical elements.

The handling of data of equivalent width time series of spectral lines of various elements, based on spectrograms of very different quality and derived by different techniques, is very demanding and time-consuming. However, well-homogenized data sometimes bear crucial information on the behavior of the phase function in the past and sometimes also in the future\footnote{It was also in the case of \ori, where the spectroscopic data helped to reveal that the spin-down is slowing and would alternate into a spin-up \citep{mik901}. The present data analysis confirms that this predicted break down moment came to pass in 2009  \citep[see][and Sect.\,\ref{901ori}]{mik11}.}.

Phase curves of equivalent widths of a particular spectral line use to be a smooth, single or double wave, which we could fit by the third order harmonic polynomial, described by Eq.\,(\ref{mono}). We can also use the following time-tested model for light curves given by Eq.\,(\ref{cosh}), which can pinpoint the phases when the spectral spots pass the stellar meridian. When using measurements of various spectral lines of an ion, the situation is more complex. As a rule, the phase curves are similar, but we must consider their different intensities and unavoidable blends with lines of other elements that, in general, somewhat suppress the relative variations of a particular spectral line. Nevertheless, the following simple model can take the circumstances mentioned above into account and give acceptable results (see Fig.\,\ref{sighe})
\begin{eqnarray}
&W_j(\vartheta)=\overline{W}_j\hzav{1+A_j\,f_{\rm ion}(\vartheta)};\quad\mathrm{where} \label{spmodel}\\ &\displaystyle f_{\rm{ion}}(\vartheta)=\frac{\sum_{k=1}^3 \hzav{\beta_{2\cdot k-1}\cos(2\,\pi\,k\,\vartheta)+ \beta_{2\cdot k}\sin(2\,\pi\,k\,\vartheta)}} {\sqrt{\sum_{l=1}^{6}\beta_l^2}}, \nonumber
\end{eqnarray}
where $W_j(\vartheta)$ is the model predicted equivalent width of the $j$-th line of the studied ion for a phase function $\vartheta$. $\overline{W}_j$ is a mean value of all equivalent widths of the $j$-th line, $A_j$ is a relative amplitude of the $j$-th line, and $f_{\rm{ion}}$ is a normalized phase curve belonging to a particular ion of the chemical element (typically, He I), the coefficients of which are the same for all studied lines of the relevant ion.

\subsubsection{Times of radio-pulse peaks}

The only type of mCP stars' observations where, instead of the standard phase curves, we used the times of some phase located events are the moments of radio-pulse peaks of \vir, known as the first main sequence radio pulsar \citep[see e.\,g.][]{trigi00}. The model describing this observation is very simple; we need just one parameter -- the phase of the center of the peak $\varphi_{\rm {0r}},\ y_{\rm p}=\The(E_i)+\tilde{P}\,\varphi_{\rm {0r}}$, while the measured quantities $y_i$ are the given times of pulses.

\subsubsection{General remarks and recommendations for the modeling of phase curves}

The suitable option of a phase curve model used for the period analysis is a crucial point of variable star modeling. Usage of an inappropriate model may influence the reliability of the complete solution and lead us to faulty conclusions. The neuralgic point of our effort is the fact that typically we are forced to use very inhomogeneous measurements of various nature and quality with phase curves connected to each other only freely. Typically, the phases of extremes of light curves, magnetic field curves, and spectral variations may coincide, but need not \citep[e.\,g.][]{silva15}. We are allowed to discuss a possible correspondence among the mentioned phase curves and use them always a posteriori, never a priori of the period analysis.

It is also desirable to plan new observations of every kind (i.e. photometric, spectroscopic and spectropolarimetric ones) so that they would be obtained more or less simultaneously (in a time interval of a few years). Then we could better phase the models of all observations with the phase information in use.

We have to pay attention to the correct weighting of entered data because the employed $\chi^2$ regression requires it (see Sect.\,\ref{solution}). If we do not know the individual uncertainties of the original data in advance or if they appear suspect, we have to estimate them iteratively from the scatter of residuals $\{\Delta y_i\}$ for appropriately defined data subsets. We recommend eliminating outlier influence using their modified uncertainties by the well-established method given in \citet{mikzej}.

The models of phase curves should be tailored to the studied object, available data, and the purpose of fitting the data. In particular, the number of used free parameters should be restricted to as few as possible, but without any serious influence on the accuracy and reliability of the results. Unfortunately, the effort of using the optimal phenomenological models considerably encumbers automation of the computational process. The diversity of real phase curves of particular mCPs requires that they have to be solved individually.

\subsection{Solution of models}\label{solution}

Our knowledge of the development of the periods of the studied variable stars in time $P(t)$ is derived from the analysis of the course of the phase function $\vartheta(t)$ whose models were dealt with in detail in Sect.\,\ref{phasefunsec}. The parameters of the adopted phase function models are calculated through minimizing $\chi^2$ quantity
\begin{equation}\label{chicko}
\chi^2=\sum_{i=1}^n \szav{\frac{y_i-y_{\rm p}\hzav{\vartheta(t_i)}}{\sigma_i}}^2,
\end{equation}
where $\{t_i,\,y_i,\,\sigma_i\}$ is the set of all available $n$ observations of phase dependant quantities $y_i$ with uncertainties (or modified uncertainties) $\sigma_i$, obtained at $t_i$. $y_{\rm p}\hzav{\vartheta(t_i)}$ is then a model prediction calculated for the particular phase function $\vartheta(t_i)$ at $t_i$. The possible models of photometric, spectropolarimetric and spectroscopic phase curves are briefly described in Sect.\,\ref{phasecurves}. We standardly assume that the shapes of phase curves are invariable (functions of the phase $\varphi(t)$, only), nevertheless, in justified cases, we can assume their slow progress in time (see Sect.\,\ref{bscirsec}).

The minimization of $\chi^2$ brings $g$ non-linear equations for $g$ unknown free parameters of all used phase curves and the phase function\footnote{The number of free parameters $g$ depends mainly on the number of 'material constants' describing all the phase curves, which could be rather large. For example, for \vir\ with $n=18\,641$ observations we need $g=274$ free parameters, whereas only four of them are necessary for the determination of the phase function. For details, see Sect.\,\ref{cuvir}.}. The equations are solved simultaneously by the standard iterative Newton-Raphson method, described e.\,g. in \citet{press,hart,mik11}, among others.

With a good initial estimate of the parameter vectors, the iterations converge fairly quickly. All estimates of uncertainties of model parameters were computed using the general law of uncertainty propagation also assuming correlations amongst individual coefficients \citep[see e.\,g.][]{bevington,mikzej}. For models of phase functions of all the studied mCP stars, except for \vir\, which exhibits cyclic changes of its period, we preferentially used the orthogonal polynomials (see Sect.\,\ref{ortopol}). This allows us to use a simple law of error propagations for evaluating functions of orthogonal polynomial parameters $\tilde{M}_0,\,\tilde{P},\,\tilde{P}',\,\tilde{P}''$ (see Eq.\,(\ref{thetaortog})).

\section{Magnetic CP stars without detectable period changes}
\begin{figure}
\centering\includegraphics[width=6cm,angle=0]{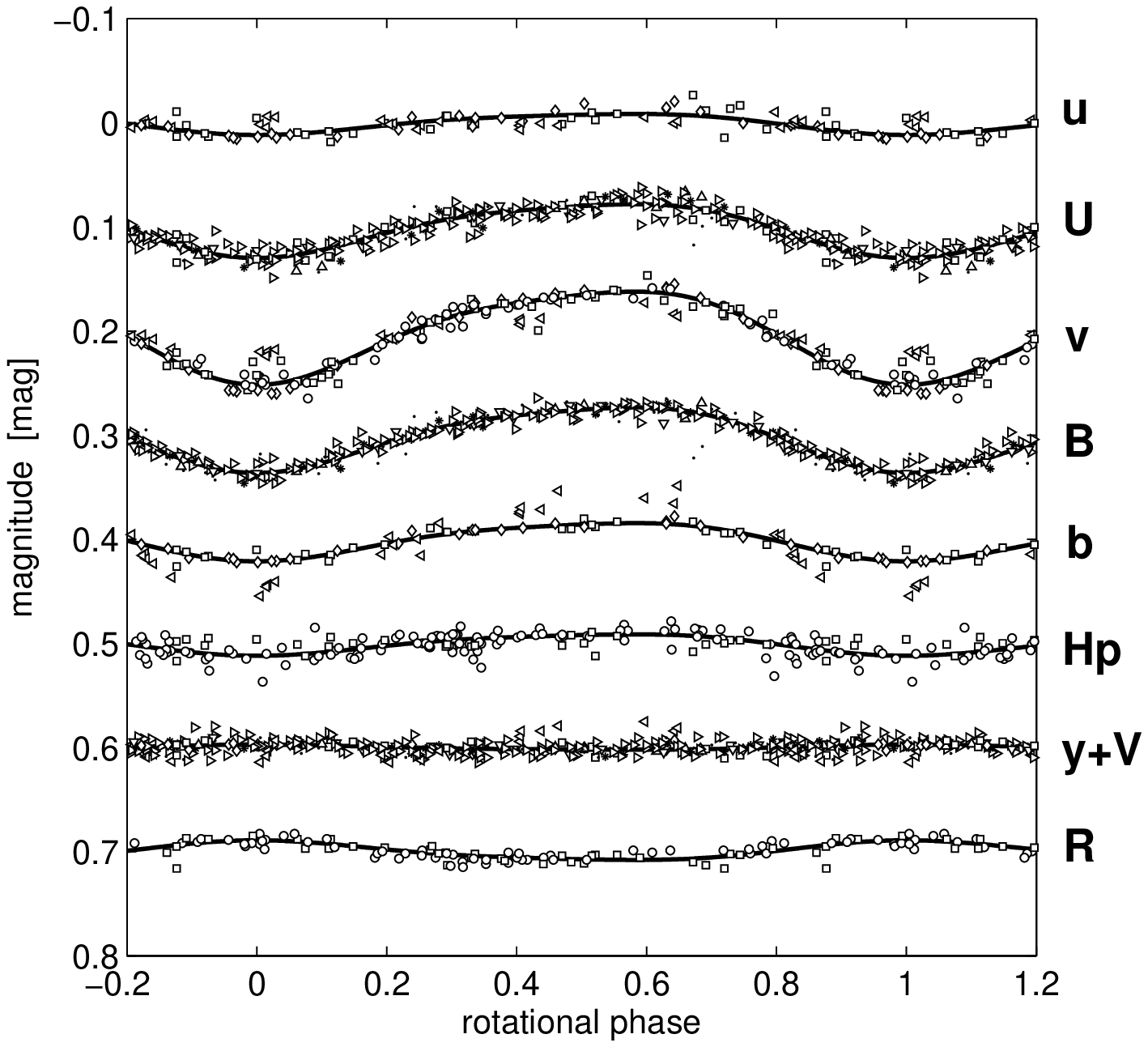}
\includegraphics[width=6cm,angle=0]{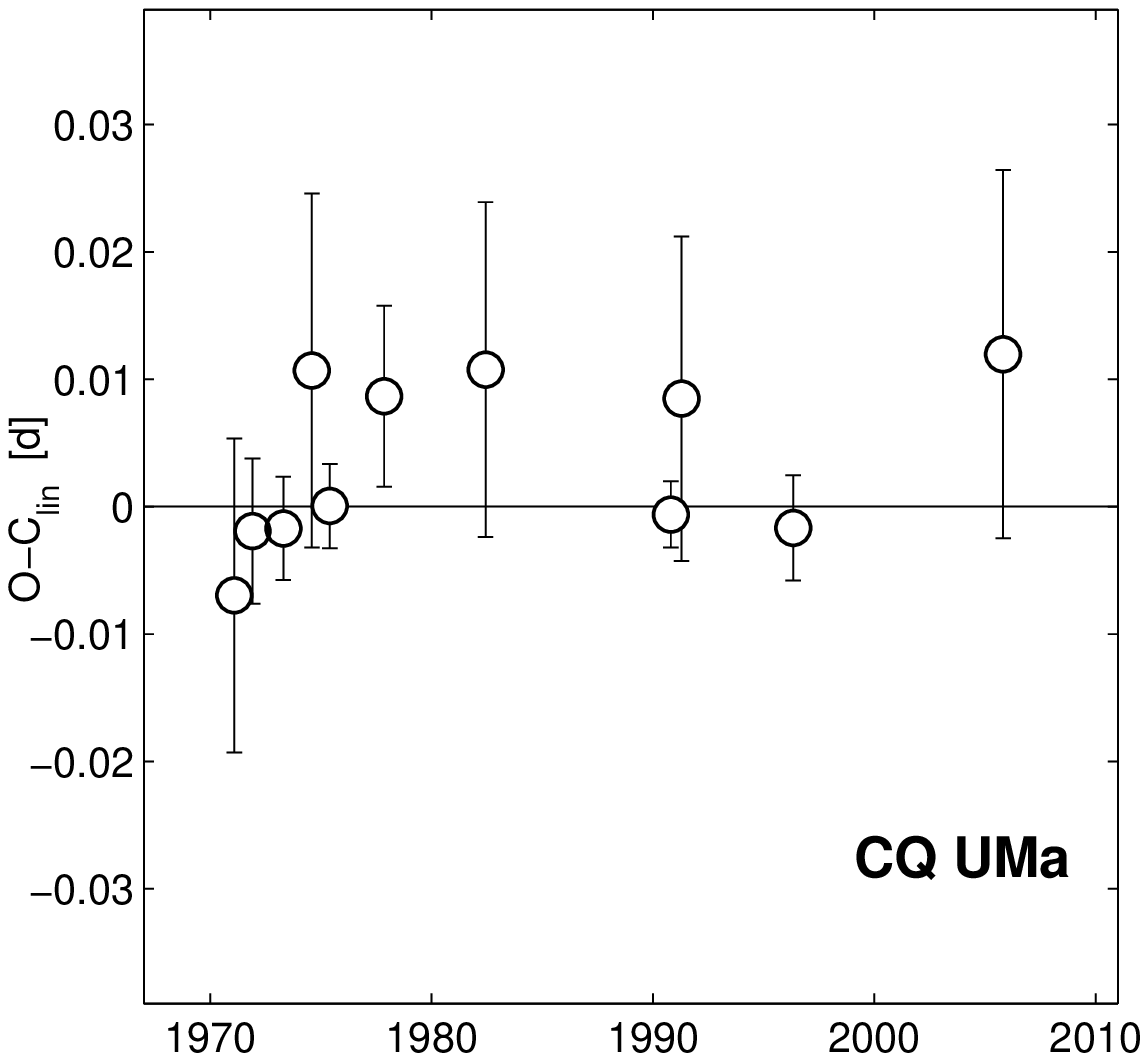}
\caption{\small (a)~CQ\,UMa light curves in $\mathit{u,\ U,\ v, B,\ b,\
Hp,\ V\!+\!y}$ and $R$--bands. Note the disappearance of
variations in $V\!+\!y$ and the antiphase variations in the
$R$--band. (b)~The time development of the
difference between the observed (O) and calculated (C) times of
the zero phase. No trend in the diagram indicates that
the rotational period of the star is constant over more than four
decades.}\label{CQkrivky}
\end{figure}

As an example of a well observed strictly periodic star, we mention CQ\,UMa~= HR\,5153~= HD\,119213. This SrCrEu Ap star displays prominent variation in the Str\"omgren $v$ band with antiphase changes in \emph{R} and \emph{I} bands (see Fig.\,\ref{CQkrivky}a).

\citet{mikper}, and \citet{unstead} used 1365 observations collected from eleven various sources of photometric data that cover a time interval of 42 years (6262 revolutions of the star). The mean period: $P\!=\!2\fd449\,912\,0(27)$ can then be derived with the accuracy of 0.23\,s.  The time derivative of the period is $\dot{P}=1(2)\times 10^{-9}=(3\pm7)$\,s\,cen$^{-1}$,
which means that period is stable, as for most other CP stars.

\citet{mikic} also tested the constancy of period of the photometrically revealed mCP star KIC 6\,950\,556, analyzing 64\,793 detrended Kepler observations of an accuracy of 0.13 mmag. They found the rotational period of the star $P_0 = 1\fd511\,785\,08(4)$\,d and the time derivative of the period $\dot{P}=1(2)\times 10^{-10}=(0.3\pm0.6)$\,s\,cen$^{-1}$. We shall note that the accuracy of the period rate determination of $2 \times 10^{-10}$ is quite sufficient to reveal all presently known mCP stars with variable periods.

Wanting to obtain the relevant statistics of the incidence of period unstable stars among the mCP stars population, we will accordingly test all known and suspected mCP stars observed by the Kepler mission.

\section{Magnetic CP stars with proven period variations}

A few mCP stars might display minor secular changes in the shape of their light curves \citep[see e.\,g.][]{zigasx}, which can be attributed to the precession of magnetically distorted stars \citep{shoade,pyper04}. However, there is also a small subgroup of mCP stars that have phase stable curves, but exhibit variable rotation periods \citep[][and references therein]{unstead,mikmos,miknat}. The constancy of their light curves on the scale of decades disqualify precession as the cause of the observed period changes \citep{mik901}.

The following text is based mainly on the period analyses of four of the best-monitored mCP stars -- \vir, \ori, \sig, and \cir, known for their period variability.

\subsection{Orthogonal polynomial model of the phase function}\label{ortogpolmodel}

As we want to compare and discuss the period development in our four mCP stars, we will use the same model for the period monitoring of all objects. It seems that the most illustrative and versatile model is the set of orthogonalized Mclaurin polynomial (Sect.\,\ref{maclaurinpol}) up to a fifth order if needed. Then, we can write
\begin{eqnarray}\label{thetaortog}
&\displaystyle \tilde{\vartheta_0}=\frac{t-\tilde{M}_0}{\tilde{P}};\quad \Delta(\tilde{\vartheta_0})=\frac{\tilde{P}'}{2}\,\theta_2+ \frac{\tilde{P}\tilde{P}''}{3!}\,\theta_3+
\frac{\tilde{P}^2\tilde{P}'''}{4!}\,\theta_4+
\frac{\tilde{P}^3\tilde{P}''''}{5!}\,\theta_5; \\
&\vartheta(\tilde{\vartheta_0})=\tilde{\vartheta_0} -\Delta;\quad\tilde{\vartheta_0}(\vartheta) \doteq \vartheta+\Delta;\quad \The(E)\doteq \tilde{M}_0+\tilde{P}\,[E+\Delta(E)];\nonumber\\
&\displaystyle P(\tilde{\vartheta_0})=\tilde{P}+\frac{\tilde{P}\tilde{P}'}{2}\frac{\rm d\theta_2}{\rm d\tilde{\vartheta_0}}+\frac{\tilde{P}^2\tilde{P}''}{3!}\frac{\rm d\theta_3}{\rm d\tilde{\vartheta_0}}+\frac{\tilde{P}^3\tilde{P}'''}{4!}\frac{\rm d\theta_4}{\rm d\tilde{\vartheta_0}}+\frac{\tilde{P}^4\tilde{P}''''}{5!}\frac{\rm d\theta_5}{\rm d\tilde{\vartheta_0}}, \nonumber\\
&\displaystyle \dot{P}(\tilde{\vartheta_0})\!=\!\tilde{P}'\!+\frac{\tilde{P}\tilde{P}''\!}{3!}\frac{\rm d^2\theta_3}{\rm d\tilde{\vartheta_0}^2}+\frac{\tilde{P}^2\tilde{P}'''\!}{4!}\frac{\rm d^2\theta_4}{\rm d\tilde{\vartheta_0}^2}+\frac{\tilde{P}^3\tilde{P}''''}{5!}\frac{\rm d^2\theta_5}{\rm d\tilde{\vartheta_0}^2},\quad \frac{\mathrm d^4\!P}{\mathrm dt^4}=P'''', \nonumber\\
&\displaystyle \ddot{P}(\tilde{\vartheta_0})\!=\!P''\!+ \!\frac{\tilde{P}\tilde{P}'''\!}{4!} \frac{\rm d^3\theta_4}{\rm d\tilde{\vartheta_0}^3}+\frac{\tilde{P}^2\tilde{P}''''}{5!}\frac{\rm d^3\theta_5}{\rm d\tilde{\vartheta_0}^3},\quad\frac{\mathrm d^3\!P}{\mathrm dt^3}=P'''+\frac{\tilde{P}^2\tilde{P}''''}{5!}\frac{\rm d^4\theta_5}{\rm d\tilde{\vartheta_0}^4},\nonumber
\end{eqnarray}
where $\tilde{M}_0,\,\tilde{P},\,\tilde{P}',\,\tilde{P}'',\tilde{P}''', \tilde{P}''''$ are parameters of the orthogonal polynomial fit which have a similar meaning as the analogous parameters $M_0,\,P_0,\,\dot{P}_0,\,\ddot{P}_0$, $P_0'''$, and $P_0''''$, introduced in the standard Mclaurin polynomial model (see Sect.\,\ref{maclaurinpol}). Nevertheless, they are not generally equal, as they have another meaning. While the parameters of the standard model are the parameters of the polynomial expansion at $t=M_0$, which is also the initial epoch in our counting system, this means that the real phase function $\vartheta$ is going through this origin and coincides with its  Mclaurin model in the close vicinity of the origin. The time-like function $\vartheta_0$ then represents the tangent of the phase function at $t=M_0$. The orthogonal model parameters $\tilde{M}_0,\,\tilde{P},\,\tilde{P}',\,\tilde{P}'',\tilde{P}''', \tilde{P}''''$ refer to the whole time interval covered by data regarding their particular weights. Thus, they are something like the mean values of the standard parameters. The auxiliary time-like function $\tilde{\vartheta_0}$ is in our orthogonal model the first (linear) approximation of the real phase function $\vartheta$. This means that it generally does not pass through the origin at the origin $t=M_0$, as well as its parameter $\tilde{P}$ is not always equal to the instantaneous period at the origin $P_0$. Therefore, the auxiliary functions $\tilde{\vartheta_0}$ and $\vartheta_0$ need not to be interchanged.

The six functions of $\tilde{\vartheta_0}$ from the set: $\{\theta_0,\ \theta_1,\ldots,\,\theta_5 \}$ are mutually orthogonal on the set of the observational data regarding their weights. So, it is valid that $\overline{\theta_j\,\theta_k}=0$ for each non-equal pair of $\theta_j$ functions.
\begin{figure}
\centering\includegraphics[width=0.90\textwidth,clip=]{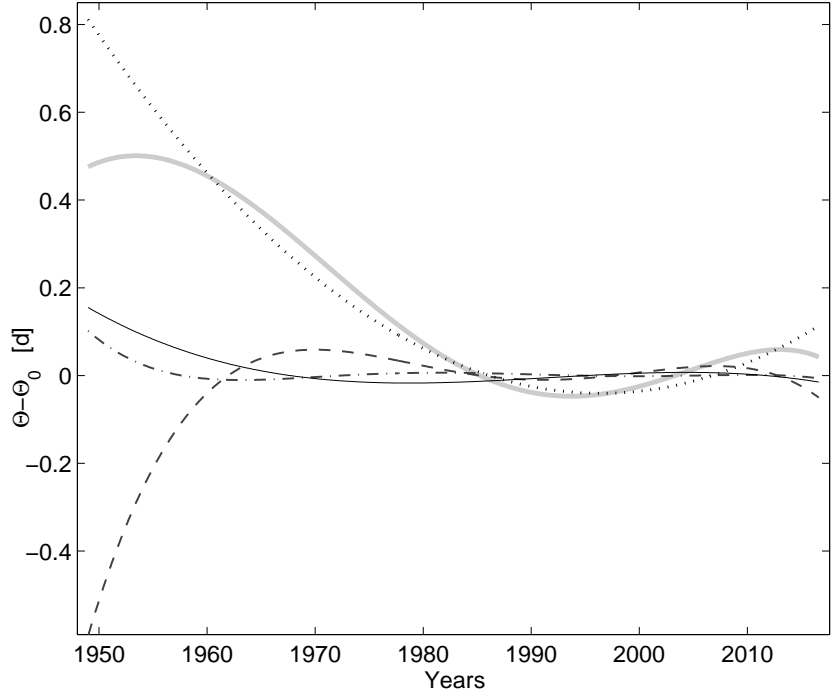}
\caption{\small The resulting fit of the observed \oc diagram ($\The\!-\!\The_0$, in days) of \vir\ by the orthogonal polynomials up to the fifth order (solid shadow line). The fit confirms the cyclic nature of the period variations observed in \vir. The contributions of the square, cubic, quartic and quintic components are marked by dotted, narrow solid, dashed and dash-dotted lines, respectively.}\label{slozky}
\end{figure}

The functions were created by the Gram-Schmidt procedure (discussed in Sect.\,\ref{ortopol})\ from the base of common polynomials $\{1,\,\tilde{\vartheta_0},\,\tilde{\vartheta_0}^2,\,\tilde{\vartheta_0}^3\ldots\}$, assuming that the origin of our epoch counting for the linear approximation is chosen so that $\tilde{M}_0\cong\overline{t}\ \Rightarrow\ \alpha_{10}=0;\ \theta_1=\tilde{\vartheta_0}$. Then, we can write:
\begin{equation}\label{ortkoefdef}
\theta_j(\tilde{\vartheta_0})=\tilde{\vartheta_0}^j-\!
\sum_{k=0}^{j-1}\,\alpha_{jk}\,\theta_k, \quad \frac{\mathrm d \theta_j}{\mathrm d\tilde{\vartheta_0}}= j\,\tilde{\vartheta_0}^{j-1}\!-\!\sum_{k=0}^{j-1}\,
\alpha_{jk}\frac{\mathrm d\theta_k}{\mathrm d\tilde{\vartheta_0}},\ \ \alpha_{jk}=\frac{\overline{\tilde{\vartheta_0}^j\,\theta_k}}{\overline{\theta_k^2}}.\\
\end{equation}
It is useful to express the set of basic orthogonal functions $\{\theta_0,\ \theta_1,\ldots,\,\theta_5 \}$ as a function of $\vt$, directly
\begin{eqnarray}\label{ortbeta}
&\displaystyle \theta_j(\vt)=\vt^j-\!\sum_{k=0}^{j-1}\,\beta_{jk}\,\vt^k,\quad \Rightarrow \quad \theta_2=\vt^2-\beta_{21}\vt-\beta_{20}, \\
&\theta_3\!=\!\vt^3\!-\!\beta_{32}\vt^2\!-\!\beta_{31}\vt\!-\!\beta_{30},\quad \theta_4\!=\!\vt^4\!-\!\beta_{43}\vt^3\!-\!\beta_{42}\vt^2\!-\!\beta_{41}\vt\!-\! \beta_{40},\nonumber\\
& \theta_5=\vt^5-\beta_{54}\vt^4-\beta_{53}\vt^3-\beta_{52}\vt^2-
\beta_{51}\vt-\beta_{50}.\nonumber
\end{eqnarray}
The computation of derivatives and the derivatives of $\theta_j$ is then trivial. Coefficients $\beta_{jk}$ are given by relations
\begin{eqnarray}\label{vypocetbeta}
&\beta_{20}=\alpha_{20},\ \beta_{21}=\alpha_{21},\ \beta_{30}=\alpha_{30}-\alpha_{32}\alpha_{20},\  \beta_{31}=\alpha_{31}-\alpha_{32}\alpha_{21},\  \beta_{32}=\alpha_{32}, \nonumber\\
&\beta_{40}=\alpha_{40}-\alpha_{42}\alpha_{20}-\alpha_{43}\alpha_{30}+ \alpha_{43}\alpha_{32}\alpha_{20},\quad \beta_{42}= \alpha_{42}-\alpha_{43}\alpha_{32}\nonumber\\
&\quad \beta_{41}=\alpha_{41}-\alpha_{42}\alpha_{21}-\alpha_{43}\alpha_{31}+ \alpha_{43}\alpha_{32}\alpha_{21},\quad \beta_{43}=\alpha_{43},\\
&\beta_{50}\!=\!\alpha_{50}\!-\!\alpha_{54}\alpha_{40}\!-\!
\alpha_{53}\alpha_{30}\!-\!\alpha_{52}\alpha_{20}\!+\!
\alpha_{54}\alpha_{43}\alpha_{30}\!+\!\alpha_{54}\alpha_{42}\alpha_{20}\!-\!
\alpha_{54}\alpha_{43}\alpha_{32}\alpha_{20}, \nonumber\\
&\beta_{51}=\alpha_{51}-\alpha_{54}\alpha_{41}-\alpha_{53}\alpha_{31}- \alpha_{52}\alpha_{21}+\alpha_{54}\alpha_{43}\alpha_{31}+ \alpha_{54}\alpha_{42}\alpha_{21}+\nonumber\\
&\alpha_{53}\alpha_{32}\alpha_{21}- \alpha_{54}\alpha_{43}\alpha_{32}\alpha_{21},\quad \beta_{53}=\alpha_{53}- \alpha_{54}\alpha_{43},\nonumber\\
&\quad \beta_{52}=\alpha_{52}-\alpha_{53}\alpha_{32}-\alpha_{54}\alpha_{42}+ \alpha_{54}\alpha_{43}\alpha_{32},\quad\beta_{54}=\alpha_{54}.\nonumber
\end{eqnarray}
The numerical results and values of pertinent dimensionless coefficients $\beta_{jk}$ of the orthogonalization for particular stars are given in Table\,\ref{tab}.

It may be worthwhile to transform parameters of the orthogonal polynomials back to the standard Mclaurin parameters, here for the polynomials of the fifth order. After some algebra, we obtain the following transforming relations
\begin{eqnarray}\label{trafo}
&\displaystyle M_0=\The(0)=\tilde{M}_0-\frac{\tilde{P}\tilde{P}'}{2}\beta_{20}- \frac{\tilde{P}^2\tilde{P}''}{6}\beta_{30}-
\frac{\tilde{P}^3\tilde{P}'''}{24}\beta_{40}-
\frac{\tilde{P}^4\tilde{P}''''}{120}\beta_{50},\\
&\displaystyle P_0=P(0)=\tilde{P}-\frac{\tilde{P}\tilde{P}'}{2}\beta_{21}- \frac{\tilde{P}^2\tilde{P}''}{6}\beta_{31}-
\frac{\tilde{P}^3\tilde{P}'''}{24}\beta_{41}-
\frac{\tilde{P}^4\tilde{P}''''}{120}\beta_{51},\nonumber\\
&\displaystyle \dot{P}_0=\dot{P}(0)=\tilde{P}'-
\frac{\tilde{P}\tilde{P}''}{3}\beta_{32}-
\frac{\tilde{P}^2\tilde{P}'''}{12}\beta_{42}-
\frac{\tilde{P}^3\tilde{P}''''}{60}\beta_{52},\nonumber\\
&\displaystyle \ddot{P}_0=\ddot{P}(0)=\tilde{P}''-
\frac{\tilde{P}\tilde{P}'''}{4}\beta_{43}-
\frac{\tilde{P}^2\tilde{P}''''}{20}\beta_{53},\quad
P_0''''=P''''(0)=P'''',\nonumber\\
&\displaystyle P_0'''=P'''(0)=\tilde{P}'''-\frac{\tilde{P}\tilde{P}''''}{5}\beta_{54}.\nonumber
\end{eqnarray}

\subsection{CU\,Virginis -- a silicon mCP star }\label{cuvir}

The famous very fast-rotating silicon mCP star \vir\ (HD\,124224,  HR\,5313) displays an intriguing period variation. It is a common hot Si-type mCP star with a mass 3\,M$_{\odot}$ and a radius of 2\,R$_{\odot}$ \citep{step} and \teff$=13\,000$\,K, $\log g=4.0,\ v\,\sin i=160$\,km\,s$^{-1},\ i=30^o$, \citep{kusch}. Its nearly dipolar magnetic field with the moderate pole strength of $B_{\rm{p}}=3.0$~kG is tilted towards the rotational axis by $\beta=74^{\circ}$, the axis inclination being $i\simeq43^{\circ}$ \citep{trigi00}. \vir\ is the first known main sequence star that shows variable radio emission, resembling the radio lighthouse of pulsars \citep{trigi,trigi11,ravi}. \vir\ also exhibits variations in light and intensities of spectral lines of He\,I, Si\,II, H\,I, and other ions. The nature of its variability in UV and optical regions has been studied by \citet{krtcu}. \vir\ now belongs to the most frequently and broadly studied mCP stars.

Occasional rapid increases in its rotation period were reported and discussed several times. \citet{pyper97,pyper98} discovered an abrupt increase of the period from $0\fd5206778$ to $0\fd52070854$ that occurred approximately in 1984 and \citet{pyper04} then discussed two possible scenarios of explanation of the observed O-C diagram, namely a continually changing period or two invariable periods. After 1998, \citet{trigi,trigi11} observed another increase in the period of radio pulses of $\Delta P=1.12$~s by comparing it with the period determined by \citet{pyper98}. \citet{pyper13} presented a very deep study updating the period development of the star based on a significant amount of excellent observational data, namely their precise 2820 Str\"{o}mgren $\mathit{uvby}$ values obtained from the Four College Automated Photometric Telescope (FCAPT) in the period 1998-2012. Their main result was that the O-C data since 1993 are consistent with a constant period of 0\fd5207137, which is the longest of all periods referred in \citet{pyper97,pyper98,pyper04}.

This statement agrees with findings of \citet{mik11} who collected and analyzed all available observations of \vir\ containing phase information between 1949 and 2011. They proved that the shapes of all phase curves were constant during several decades, while the period was continually changing. The rotation period was gradually shortening until the year 1968, when it reached its minimum. The period then started increasing, reaching its local maximum. The following spin-up was recently undoubtedly confirmed by \citet{krtosc}.

The present period analysis is also based on published FCAPT photometry from 1998-2012 by \cite{pyper13} and our measurements from 2011-6. The complete observational material represents now 19\,641 individual measurements of \vir\ including photometric measurements in photometric bands from 200 to 753 nm \citep[see in][]{krtcu}, as well as spectroscopic, spectropolarimetric and radiometric observations.

\begin{figure}
\centering\includegraphics[width=0.99\textwidth]{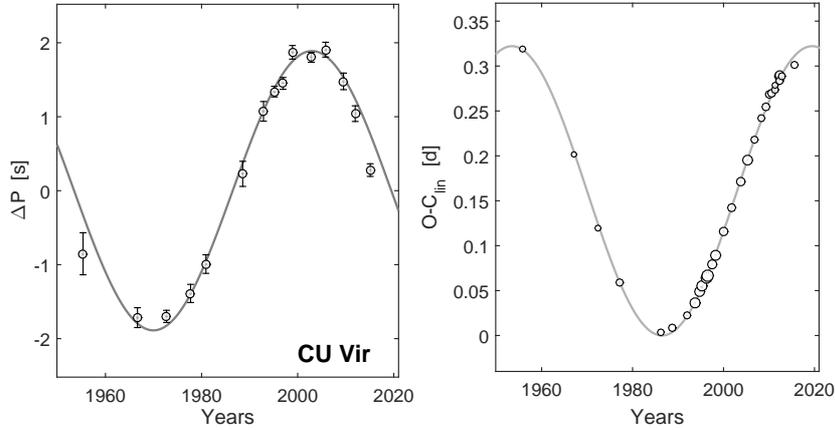}
\caption{\small (a) Changes in the rotation period $\Delta P=P(t)-P_0$ of \vir\ in seconds with respect to the mean rotation period $P_0$. The period variations are approximated by a sinusoid reaching its extrema in 1970 and 2003 (see Eq.\,(\ref{cuper})). The amplitude of period variations is 3.78\,s. (b) O-C$_{\rm{lin}}=\The\!-\!\The_0$ variations have a semiamplitude of $0.1611(5)$\,d\,$=0.31\,\overline{P}$ (see Eq.\,(\ref{cuth})).}\label{duoCU}
\end{figure}
Let us assume, in accordance with \citet{krtosc}, that the difference $\Delta(\phi)$ is a sinusoidal function of $\phi(t)=(t-T_0)/\mathit{\Pi}$, where $T_0$ is the origin of counting of cycles with the period $\mathit{\Pi}$. Applying relations (\ref{cycpsi}), we obtain
\begin{eqnarray}
&\displaystyle \Delta=\frac{A}{P_0}\hzav{1\!-\!\cos\zav{2\,\pi \phi}};\quad \vartheta_0=\frac{t-M_0}{P_0};\quad \phi(\vartheta_0)= \frac{P_0\,\vartheta_0+M_0-T_0}{\mathit{\Pi}}; \nonumber\\
&\vartheta=\vartheta_0-\Delta(\tilde{\vartheta_0});\quad \The_0(E)=M_0+P_0 E;\quad \The(E)=\The_0(E)+P_0\Delta(E);\label{cuth}\\
&\displaystyle P(\vartheta_0)=P_0+\frac{P_0^2}{\mathit {\Pi}}\frac{\mathrm d\Delta}{\mathrm d\phi}=P_0\hzav{1+\frac{2\,\pi A}{\mathit{\Pi}}\,\sin\zav{2\,\pi\,\phi}},\label{cuper}
\end{eqnarray}
where $A$ is a semiamplitude of the change $\The(E)-\The_0(E)$ with the minimum at $T_0$, the semiamplitude of the mean period undulation being $A_{\mathrm P}=2\,\pi AP_0/\mathit{\Pi}$. $M_0$ was chosen so that $\Delta(\tilde{\vartheta_0}=0)=\mathrm d\Delta/\mathrm d \vartheta_0=0$. Analysing all the available observational data of \cir, we found $M_0=2\,446\,604.4390$ (fixed), $P_0=0.520\,694\,04(3)\,\mathrm d,\ T_0=2\,446\,604(13),\ \mathit{\Pi}=24\,110(150)\,\mathrm d=66.0\pm0.4$\,yr, $A=0.1611(5)$\,d, and $A_{\mathrm P}=1.888$\,s (see Fig. \ref{duoCU}). The employed data cover more than one cycle of the proposed sinusoidal variations.

The most natural explanation for the \vir\ O-C diagram offers the light-time effect. This, however was refused by \citet{pyper97,pyper13}, and
Mi\-ku\-l\'a\-\v{s}ek et al. (2011a).
Therefore  we have to find a different explanation, admitting the fact that the observed period variations are due to changes in the angular velocity of the star or its surface layers at least \citep[see also][]{step}.

A very ambitious explanation of the observed behavior of \vir\ was proposed by \citet{krtosc}, introducing a novel mechanism of rotation oscillation as a consequence of internal waves spreading within a rotating magnetic star.

We can also describe the observed phase function of \vir\ by the orthogonal polynomial model of the fifth order (see Sect.\,\ref{ortogpolmodel}), determined by the set of coefficients $\beta_{jk}$, given by Eq.\,(\ref{ortkoefdef}) and (\ref{vypocetbeta}) and 6 parameters of the model (all these results necessary for ephemeris calculations are listed in Table \ref{tab}). $\tilde{M}_0=2\,451\,217.2306$, the mean weighted period is $\tilde{P}=0\fd520\,700\,95(15)$. The parameters $\tilde{P}',\ \tilde{P}'',\ \tilde{P}'''$ and $\tilde{P}''''$ are determined with good accuracy, only the last parameter, $\tilde{P}''''=8(3)\times10^{-24}$\,d$^{-4}$, is so uncertain that we can neglect it. Moreover, it influences only the very beginning of \vir\ measurements, which at the time were rather unreliable. Fig.\,\ref{slozky} displays the contributions of particular terms of the orthogonal polynomial model of a high degree. Nevertheless, it seems that the validity of the sinusoidal model is more or less confirmed, even if we used the model not assuming any cyclicity of period variations.

\subsection{V901 Orionis -- a He-strong mCP star }\label{901ori}

V901 Orionis = HD 37\,776 is a very young He-strong B2p mCP star with an effective temperature of 23\,000 K \citep{cid}, residing in the emission nebula IC 432, with a global, extraordinarily strong $(B_s\approx20$~kG), and complex magnetic field \citep{thomla,ko901}. The observed moderate light variations are caused by the spots of overabundant silicon and helium \citep{krt901}.

Thirty years of accurate photometric and spectroscopic monitoring enabled us to reveal a continuous rotational deceleration
(Mikul\'a\v{s}ek et al., 2007a; Mi\-ku\-l\'a\-\v{s}ek et al., 2008),
increasing the period of about $1\fd5387$ by a remarkable 18 s! Ruling out (a) a light-time effect in a binary star, (b) the precession of the star's rotational axis, and (c) evolutionary effects as possible causes of the period change, we interpreted the deceleration in terms of the rotational braking of the outer stellar layers as caused by the angular momentum loss in the stellar magnetosphere.

However, this cannot explain the discrepancy between the spin-down time, $\tau = P/\dot{P}=3 \times 10^5$ yr, and the star's age of one million years, or older \citep{mik901,unstead,mik11}. The interpretation of the rotation period evolution by a simple angular momentum loss was also questioned by the negative value of the second derivative of the period: $\ddot{P}=-29(13)\times 10^{-13}\,\mathrm{d}^{-1}$, which indicated that braking could soon change into acceleration \citep{mik901}. New precise measurements prove this conclusion without any doubts.

\begin{figure}
\centering\includegraphics[width=0.99\textwidth]{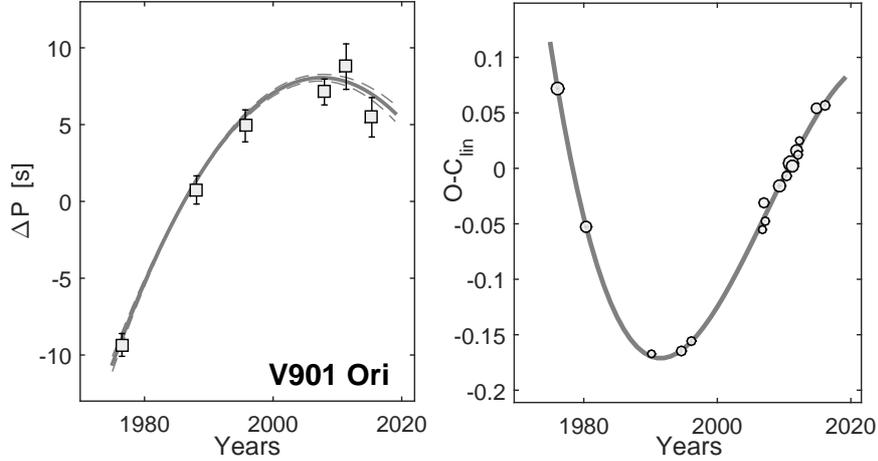}
\caption{\small (a) Changes in the rotational period, $\Delta P= P(t)-\tilde{P}$, of \ori\ in seconds. Time dependence of the period is approximated by a parabola reaching its maximum in 2009 (see Eqs.\,(\ref{thetaortog}) and (\ref{ortbeta})). (b) Changes in times of the zero phase in days versus the linear model, O-C$_{\rm{lin}}=\The\!-\!\The_0$, can well be fitted by a cubic parabola (see Eq.\,(\ref{901vt})).}\label{duo901}
\end{figure}
Presently, we have at our disposal 3656 photometric, 663 spectroscopic, and 75 magnetic measurements covering more or less evenly a time interval of 40 years. All the data were simultaneously modeled with the quartic (fourth order) orthogonal polynomial phase function model $\vartheta(\vt)$, described in Sect.\,\ref{ortogpolmodel}
\begin{equation}\label{quartic}
 \vt=\frac{t-\tilde{M}_0}{\tilde{P}};\quad  \Delta=\frac{1}{2}\,\tilde{P}'\,\theta_2+ \frac{1}{6}\,\tilde{P}\tilde{P}''\,\theta_3+ \frac{1}{24}\,\tilde{P}^2\tilde{P}'''\, \theta_4,
\end{equation}
where the parameters of the model $\tilde{M}_0,\ \tilde{P},\ \tilde{P}',$ $\tilde{P}''$, and $\tilde{P}'''$ are listed in Table\,\ref{tab}, including coefficients $\beta_{jk}$, necessary for ephemeris calculations.

The parameters of the orthogonal model are $\tilde{M}_0=2\,453\,348.710(3)$, $\tilde{P}=1\fd538\,728\,6(5)$, $\tilde{P}'=1.12(3)\times 10^{-8}$, $\tilde{P}''=-3.05(25)\times 10^{-12}\,\mathrm d^{-1}$, and $\tilde{P}'''=1(3)\times 10^{-16}\,\mathrm d^{-2}$. It seems that the last term of the expansion of the phase shift $\Delta(\vt)$ (with $\theta_4$) can now be completely neglected\footnote{We should note that the uncertainty of the quartic term is quickly and proportionally diminishing to $n^{-1/2}\Omega^{-4} \sim \Omega^{-9/2}$ [see (\ref{chyby})], where $n$ is the total number of observations and $\Omega$ is the time of observations of the star. Consequently, in 15 years of constant monitoring, it will drop to $10^{-12}$\,d$^{-2}$, which should be enough for solving the question whether the variations of the period are cyclic or not.}.

\begin{figure}
\centering\includegraphics[width=0.90\textwidth,clip=]{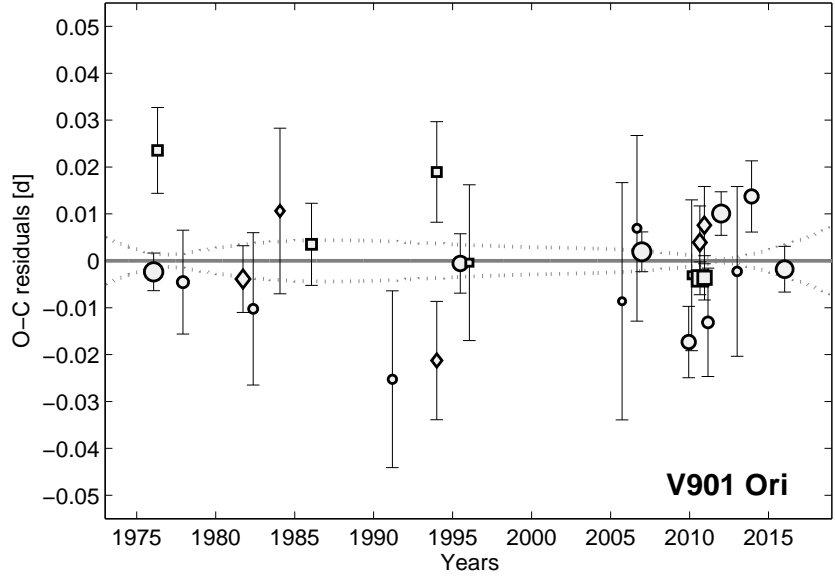}
\caption{\small O-C residuals of individual sets of observations of \ori\ that were obtained by different techniques related to various structures on the surface. Circles mark results based on photometric measurements (silicon spots) while squares and diamond signs correspond to spectropolarimetric (magnetic field geometry) and spectroscopic (helium spots) observations. The areas of markers are proportional to the weights of individual phase shift determinations. Dotted lines denote the one $\sigma$ uncertainty in the model fit.}\label{resid901}
\end{figure}

According to equations (\ref{thetaortog}) we can predict the course of the phase function $\vartheta(\vt)$ the moments of the zeroth phase $\The(E)$ and that computed by the linear approximation $\The_0(E)$ for any epoch $E$ as follows
\begin{equation}\label{901vt}
\vartheta(\vt)=\vt-\Delta(\vt),\quad \The_0(E)=\tilde{M}_0+\tilde{P} E,\quad \The(E)= \The_0(E)+\tilde{P}\Delta(E),
\end{equation}
as well as the courses of the instantaneous period $P(\vt)$ and its derivatives $\dot{P}(\vt),\ \dot{P}(\vt)$. Using them we can conclude that the deceleration of the stellar rotation switched in 2009.7(1.0) to acceleration instead (see Figs.\,\ref{duo901} and \ref{resid901}).

It is likely that the monitored part of the \oc\ curve is only a segment of a cyclic curve, but we are not yet able to determine its parameters, like as amplitude and period $\mathit\Pi$. We can only estimate that the period must be longer than observed in hundred years. Then the reasons for the period change could be the same as that of \vir, but they necessarily need not be \citep{krtosc}.

\ori\ has been followed up for the last forty years as a photometric, spectroscopic and magnetic variable, with each type of variations giving us information about the location of another type of surface structures. Photometric spots, accountable for light variations are determined namely by the distribution of silicon on the star's surface \citep{krt901}, while spectroscopic variability was studied solely using changes of the He\,I spectral line intensity \citep{mik901}. This informs us about the location of spots with an overabundant helium, which do not coincide with silicon spots \citep{choch}. Observations of the effective magnetic field open up the possibility to study the magnetic geometry of the star \citep{thomla,ko901}.

If the mentioned structures have been moving mutually during the last decades, we should detect another trend in residuals of the three types of observations. Fig.\,\ref{resid901} clearly proves that there were no such motion, so we can conclude that outer layers of the star rotate as a solid body. That fact confirmed the idea that stellar atmospheres of mCP stars with their strong global magnetic fields are horizontally stabilized due to this property.

\subsection{\sig\ - a hybrid of mCP and Be star}

\begin{figure}
\centering\includegraphics[width=0.90\textwidth,clip=,angle=0]{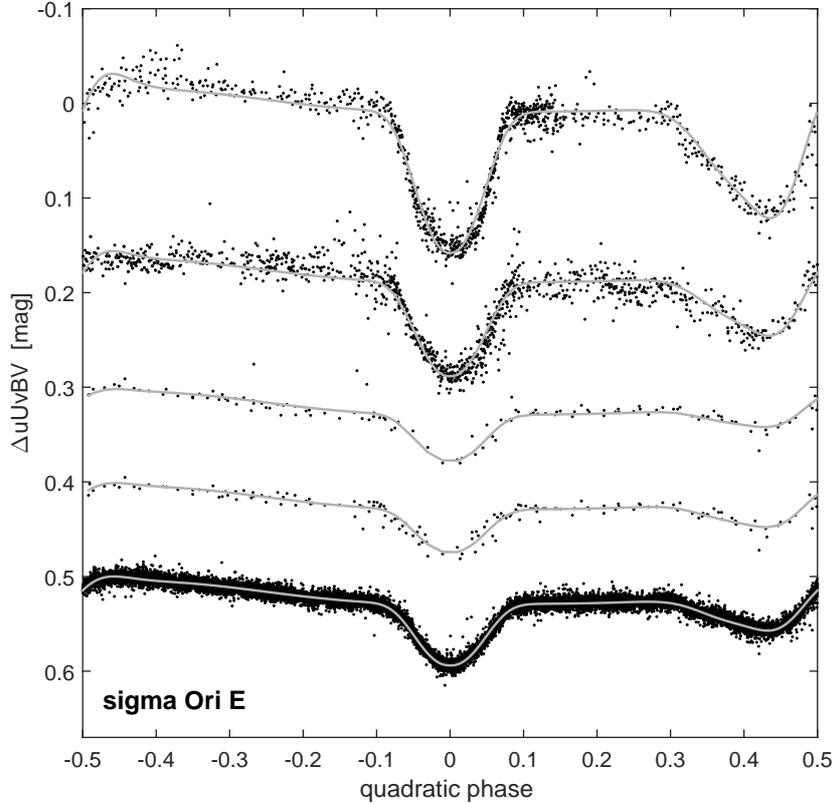}
\caption{\small $\Delta uUvBV$ light curves of \sig\ (from top to bottom) plotted versus the quadratic phase. The points are observations; solid lines denote the model fits.}\label{phsig}
\end{figure}

$\sigma$\,Ori\,E = HD 37479 = V1030 Ori is a hybrid of a classical He-strong mCP star and a Be star with strong stellar winds. It is an extremely young and massive star with a period of photometric and spectral changes $P=1\fd19801(1)$ \citep{hesser, hunger}. The spin-down of the star has been noticed by \citet{reiners} and confirmed by \citet{oxa}. \citet{town} then revealed the period to be linearly increasing at the rate $\dot{P}=2.89(21)\times 10^{-9}=0.091(7)$\,s\,yr$^{-1}$ using the method of the analysis of five times of primary minimum. The authors explained the observed lengthening of the period by magnetic braking through strong stellar winds. The established spin-down time $\tau=P/\dot{P}=1.4\times10^6$  years was considered to be compatible with the estimated age of the star \citep{unstead}.
\begin{figure}
\centering\includegraphics[width=0.90\textwidth,clip=,angle=0]{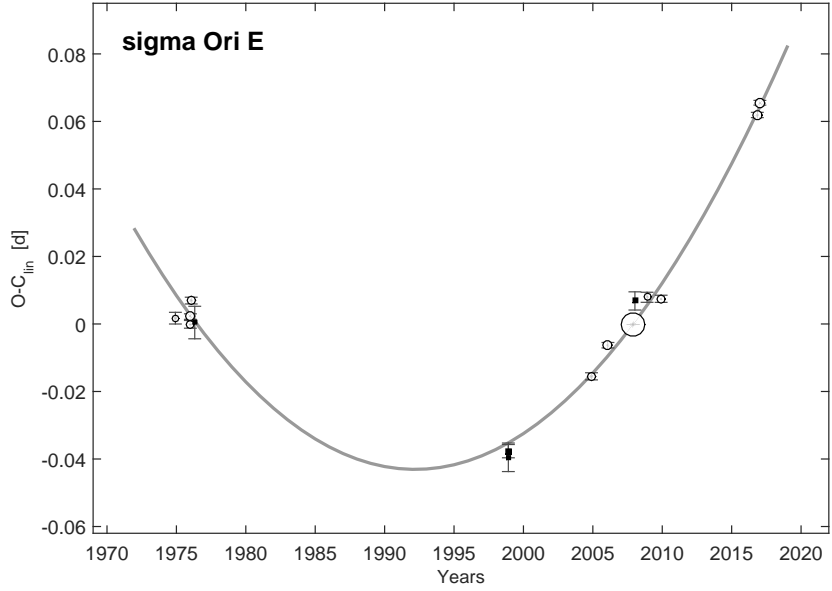}
\caption{\small The O-C$_{\rm{lin}}=\The\!-\!\The_0$ diagram of \sig\ with the parabolic fit and the mean values of individual sets of observations. Circles are based on the analysis of photometric data; black squares correspond to the analysis of helium lines' intensities. The areas of the symbols express the weights of observational sets.}\label{ocsig}
\end{figure}

We considerably broadened the extent of the observations used by \citet{town}, added observations from the MOST satellite \citep[courtesy of][]{townmost}, some archival photometric observations, and our own $u$ observations from the winter 2016/17, combining them with our own and archival observations of equivalent width measurements of strong helium lines. In total, we used 27\,656 individual measurements, 27\,373 being photometric ones, covering a time interval of 43 years (1974--2017). Light curves (see Fig.\,\ref{phsig}) were modeled by special periodic functions originally developed for eclipsing binaries \citep{mikecl}, assuming asymmetry of their minima. Using the whole mentioned material and a more sophisticated method of the phase shifts we were able to enhance the accuracy of the $\dot{P}$ determination 8-times. The full list of the used data and special models of phase curves will be published elsewhere soon.

We give here only the results of the phase function fitted by the model of the quartic orthogonal polynomial, which is formally identical with that used for \ori\ (see Eq. (\ref{quartic}) and (\ref{901vt}), and the references around). The parameters and coefficients $\beta_{jk}$ are listed in Table \ref{tab}.

We found the following model parameters: $\tilde{M}_0=2\,454\,296.5980(2)$, the mean period $\tilde{P}=1.190\,836\,98(6), \tilde{P}'=3.08(3)\times 10^{-9}=0.0981(9)$\,s\,yr$^{-1}$, and $\tilde{P}''=3(1)\times10^{-13}$\,d$^{-1}$. The points in the O-C diagram are excellently fit by a simple parabola (see Fig.\,\ref{ocsig}); the influence of a cubic term, if any, is obviously negligible. The change of the period is almost linear so that it could also be caused by mechanisms of the power law category. \sig\ is the hottest of the stars with varying periods, so its stellar winds could be dense enough to explain its moderate rotational breaking.

Up to 2010, it seemed that all rotationally unstable mCP stars belong to the hotter and thus more massive and younger mCP stars \citep{unstead}. Naturally, the observed affinity may only be the result of the wrong conclusion drawn from a very limited sample of such stars. Therefore we directed our analyses also to short-periodic moderately cool mCP stars, also called as SrCrRE or SiCr stars exhibiting light variations with a large amplitude. All studied moderately cool mCP stars turned out to be stable with one exception described below.

\begin{figure}[pth!]
\centerline{\includegraphics[width=\textwidth,clip=]{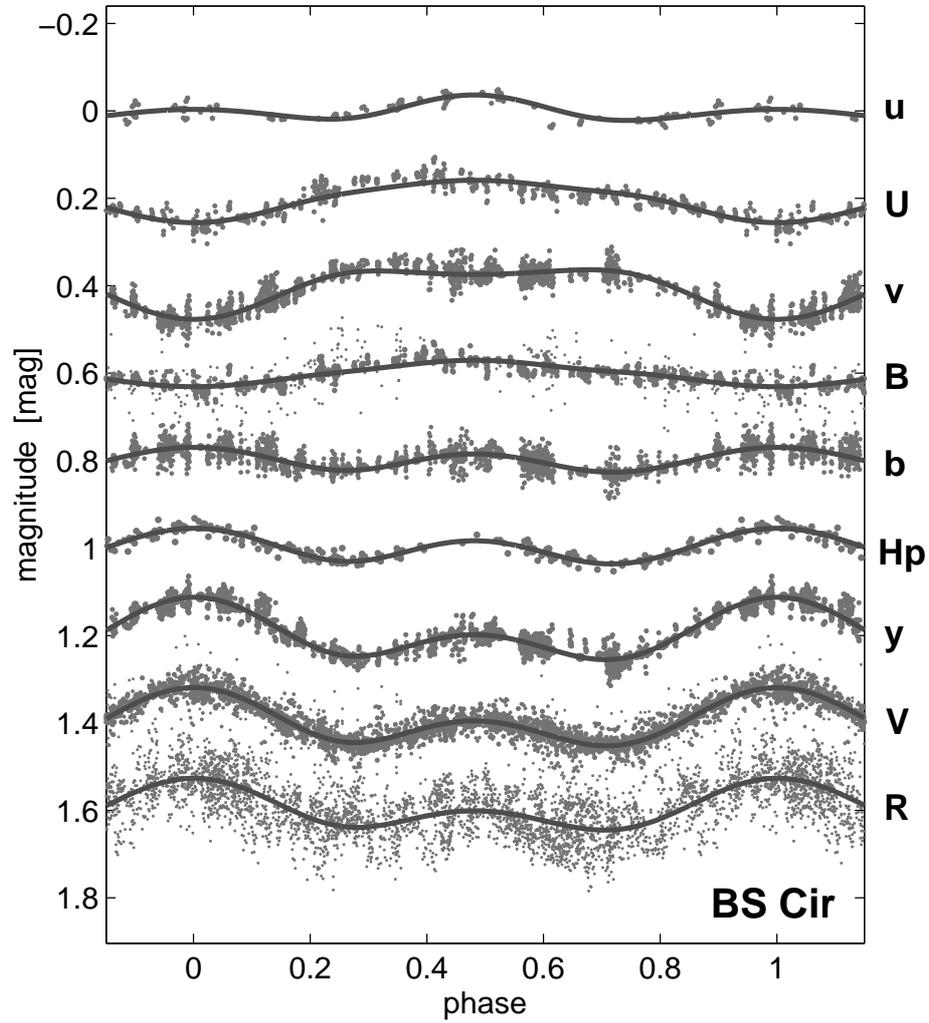}}
\vspace*{10.mm}
\caption{\small Light curves of \cir\ in various filters arranged according to their effective wavelengths. The shapes of light curves can well be interpreted by the model with two photometric spots centered at phases 0.0 and 0.48 with the contrasts depending on their effective wavelengths. The rotational phase is calculated according to the ephemeris with a quadratic term. The area of full circles is equal to their weights.} \label{BScurve}
\end{figure}

\subsection{\cir\ -- a moderately cool surprise}\label{bscirsec}
\cir\ = HD\,125630 = HIP\,70346 is a southern mCP star of the A2pSiCr type that is a representative of moderately cool magnetic chemically peculiar stars that display rather strong light variations in Str\"omgren index $c_1=(u-v)-(v-b)$ indicating large changes in the height of the Balmer jump. Combining available kinematic, photometric and spectroscopic data of the star, \citet{mikmos} derived the following astrophysical parameters: $T_{\mathrm{eff}}=8800\pm500$\,K, $L=41.7\pm1.4\,\mathrm{L}_{\odot},\ M=2.32\pm0.14\,\mathrm{M}_{\odot},\ \mathrm{age}= 510^{+90}_{-150}$\,Myr. A moderate global magnetic field with bipolar strength $B_{\rm{p}}$ of several kG is present
\citep{kobacu,hub06}.

\cir\ was observed in 1975-6 by \citet{vofa79} in $\textit{uvby}$ and then in 1980 by \citet{mare80,mare83,mama85} have revealed the star to be photometrically variable with a period $P=2\fd205\pm0\fd004$ \citep[apparently not taking into account the photometry of][]{vofa79}. A data set of \citet{mare83} was later reanalysed and the formerly found period was confirmed by \citet{mama85}. \citet{cale93} then added \citeauthor{mare83} data to their 56 precise $\mathit{uvby}$ measurements taken in 1991 and `improved' the period to $P'=2\fd20552(6)$. As they again did not consider \citeauthor{vofa79} data, they could be mistaken in the total number of cycles between 1980 and 1991 ($\Delta t \simeq 11\times 365\simeq4015$\,d) by one $(\Delta k=-1)$. The incorrect period is also given in the list of \citet{dubath} containing periods and types of variable stars in the Hipparcos survey where \cir\ was treated as an eclipsing binary with the orbital period of 1\fd1020.

\begin{figure}[h]
\centerline{\includegraphics[width=0.99\textwidth,clip=]{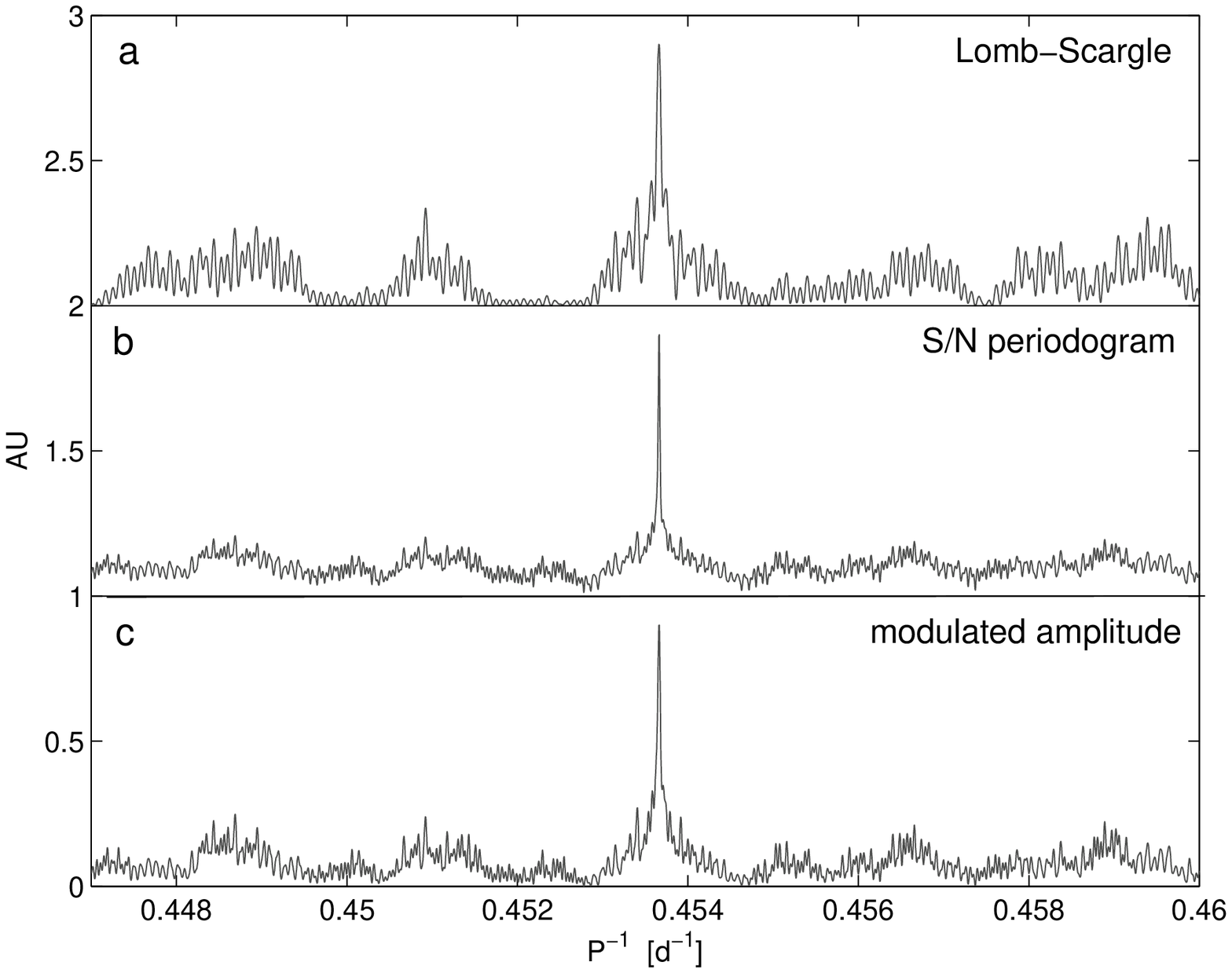}}
\begin{center}
\caption{\small All three types of various periodograms of \cir\ undoubtedly pinpoint the only dominant period peak at $P=2\fd2042$ \citep[for details,  see][]{mikper}.} \label{bsperds}
\end{center}
\vspace*{-9.mm}
\end{figure}

\begin{figure}[h]
\centerline{\includegraphics[width=0.99\textwidth,clip=]{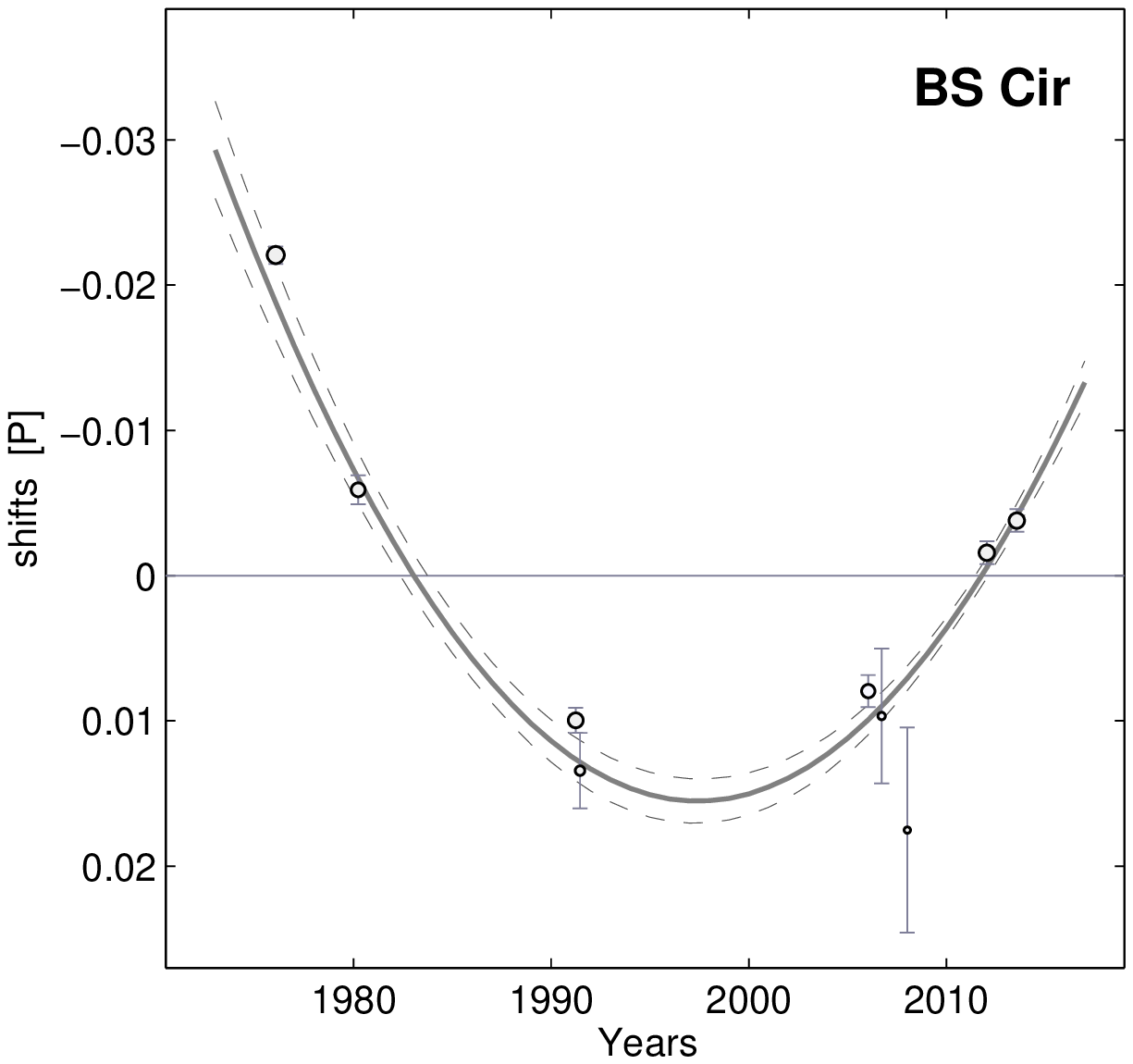}}
\begin{center}
\caption{\small The phase shifts of observed light curves versus light curves predicted by the linear ephemeris $\vartheta-\vt$, can be well fitted by a parabola. This indicates that the observed period of light variations is linearly lengthening with the rate: $\dot{P}=5.4\times10^{-9}$. The one-$\sigma$ deflections from the fit are represented by dashed lines, the areas of individual markers are inversely proportional to their uncertainty, which is also signed by error bars.} \label{ocbs}
\end{center}
\vspace*{-9.mm}
\end{figure}

The first actual \cir\ period of 2\fd2043, derived from the analysis of all the above mentioned data sets, replenished by Hipparcos, ASAS, Pi of the Sky observations and our own observations in the South Africa Astrophysical Observatory by various photometric instruments in 2011-4, was briefly reported in the review paper of \citet{mikmos}. The ephemeris of \cir, based on 14\,488 photometric measurements in eleven data sets that cover, more or less evenly, a time interval of 38 years, also contained the square term. It showed undoubtedly that the instantaneous period of the star is rising with the rate $\dot{P}=5.6(4)\times 10^{-9}=0.181(13)\,\mathrm{s\,yr}^{-1}$.

\citet{mikper} argued that periodograms of \cir, based on all the photometric material, admit the only mean period of 2\fd2042 (see Fig.\,\ref{bsperds}). The nature of the star is then more extensively dealt with in the little monograph of \citet{mik15}, based on the rich photometry and also the spectroscopy. However, the period analysis was based solely on the more reliable photometric data. Light curves obtained in {\it uvbyUBVRIHp} photometric colors were successfully fitted with the model expressed by Eq.\,(\ref{cosh}) assuming two differently colored, centrally symmetric photometric spots centered at the phases $\varphi_{02}=0.0000(5)$ and $\varphi_{02} = 0.4802(11)$, with half-widths of $d_1=0.136(2),\ d_2=0.117(3)$ (see Fig.\,\ref{BScurve}).

The different spectral energy distribution in the spots shows that at least two diverse mechanisms of the redistribution of the radiative energy are active there.
According to Eq.\,(\ref{ortkoefdef}) and (\ref{vypocetbeta}), we found coefficients for the quartic orthogonal polynomial $\beta_{jk}$, listed in Table\,\ref{tab}. Then, we derived the quartic orthogonal ephemeris for the phase function $\vartheta(\vartheta_0)$, formally identical with the same relations for \sig\ (Eq.\,(\ref{quartic})). The origin and the period of the linear approximation are: $\tilde{M}_0=2\,453\,943.3871(12)$, $\tilde{P}=2\fd204\,284\,86(7)$. The mean lengthening of the observed period of $\dot{P}=5.4(4)\times 10^{-9}=0.170(13)$\,s\,yr$^{-1}$ is well established, opposing to higher terms that can be neglected.

Currently, \cir\ belongs to the best and longest monitored mCP stars. However, we are not able to mine more information from the period analysis, because there are no signs of any elongation from the simple model assuming a steady rise of the period. Nevertheless, some information we can be gain from possible secular changes in the light curves themselves. \citet{mik15} have found small long-term trends in the shapes of the light curve (the amplitude of both spots is decreasing) while the phase distance between the centers of the spots remains constant. Such moderate changes (if real) could bear evidence of a slow free precession of the stellar body.

\section{Comparison of mCP stars having unstable periods.\\ Discussion}\label{diskuse}

The relative number of stars showing rotational period changes on the time scale of decades among single upper main sequence stars was theoretically quite unpredicted, and until now the true cause (or causes) of such peculiar behavior is not known. The main aim of the present analysis of more than 60 thousand mostly good photometric measurements of four magnetic CP stars, suspected of unstable rotation, was to prove those suspicions and find the properties of this phenomenon which could lead us to the solution of that puzzle.

The analysis confirmed without any doubts that all four stars mentioned above, namely \vir, \ori, \cir, and \sig, display changes in their observed period of variability with a certainty of more than 13 $\sigma$. As we have shown in the case of \ori\ (for details, see Sect.\,\ref{901ori}), the phase curves and their relative phase shifts remain constant during several decades, which indicates that the outer layers of mCP stars are extraordinarily stable including the magnetic geometry and the locations of photometric and spectroscopic features. This stability is very likely caused by the strong global magnetic field frozen into the atmospheric plasma that fastens the surface of the star, so it behaves like a solid body. Thus, the monitoring of the period changes tells us something about rotation of this outer, directly observable part of mCP stars. The found gradual variations in the angular velocity of several mCP stars could, but need not, inform us about the rotation of the inner part of the star. Nevertheless, the finding of the physical mechanism causing the observed rotational variations represents an interesting astrophysical challenge both for observers and theoreticians.

The studied subgroup of mCP stars consists of very disparate members with a minimum of common properties. \vir\ is a unique main-sequence radiopulsar, \ori\ is one of the hottest mCP stars exhibiting a strong and extraordinarily complex magnetic field, \sig\ is a hybrid between a Be star and a He-strong star with the very strong magnetized stellar wind, and the last star is an old, moderate cool mCP star. It seems that the only shared common property is the fact that their periods are shorter than those of the majority of magnetic chemically peculiar stars \citep{mikper}. Unfortunately, this might only be due to the selection effect following from relations in Eq.\,(\ref{chyby}), according to which uncertainties in the determination of the period and period changes are proportional to the square of the period ($\sim P^2$).

The relevance of the individual parameters of the expansion can be judged by a comparison of their numerical values and the estimates of their uncertainties, $\delta P_0,\ \delta \dot{P}_0,\ \delta{\ddot{P}_0}$

\begin{equation}\label{chyby}
q=\frac{\sigma\,P_0^2}{A\sqrt{n}};\quad \delta P_0=\frac{1.55\,q}{\Omega}; \quad \delta \dot{P}_0=\frac{12\,q}{\Omega^2}; \quad \delta \ddot{P}_0=\frac{140\,q}{\Omega^3},
\end{equation}
where $q$ is an auxiliary quantity, $\sigma$ is the mean scatter of data used for the ana-lysis, $n$ is the number of measurements, $A$ is the mean amplitude of variations, and $\Omega$ is the time interval covered by observations\footnote{Applying formula (\ref{chyby}) to V901\,Ori, one of the best photometrically monitored variables among mCP stars with parameters: $P=1.539$\,d, $\sigma=0.0048$\,mag, $A=0.040$\,mag, $\Omega=40$\,yr, $n=4500$, the uncertainties of the period $P$ and its first and second derivatives $\dot{P}$ and $\ddot{P}$, can be estimated as $\delta P=4.5\times10^{-7}$\,d\,$=0.04$\,s; $\delta\dot{P}=2.4\times10^{-10}=0.76$\,s\,cen$^{-1}$,\ $\delta \ddot{P}\simeq2\times10^{-13}$\,d$^{-1}$.}. The relations show that the best chance of revealing period changes exists in continuously monitored stars with well defined phase curves and periods as short as possible.

\begin{table}\scriptsize
\caption{The comparison of some physical characteristics of \vir, \ori, \sig, and \cir\, as mentioned in Sect.\,\ref{diskuse} and the results of their orthogonal polynomial model of the phase functions (see Sect.\,\ref{ortogpolmodel}).}\label{tab}
\begin{tabular}{l|cccc}
  \hline\hline
 Name & \vir & \ori & \sig & \cir \\
 \hline
 $T_{\rm {eff}}$ [K]&13\,400&23\,000&23\,000&8\,800\\
  sp. type&B8p, SiCr& B2p, He-strong&B2ep, He-strong &A4p, CrEu\\
  age [Myr]&90&3&1-2&510\\
  $\tilde{M}_0$ [d] & 2\,451\,217.2306(3) & 2\,453\,348.710(3)& 2454296.5980(2) & 2453941.3871(12) \\
  $\tilde{P}$ [d] & 0\fd520\,700\,95(15) & 1\fd538\,728\,6(5) & 1\fd190\,836\,98(6) & 2\fd0428486(7)\\
  $\tilde{P}'$  & $2.796(4)\times 10^{-9}$ & $1.12(3)\times 10^{-8}$ & $3.08(3)\times10^{-9}$ & $5.4(4)\times10^{-9}$ \\
  $\tilde{P}'' $[d$^{-1}$] & $-7.3(3)\!\times\!10^{-14}$ & $-3.05(25)\times 10^{-12}$ &[$3(1)\times 10^{-13}$] & [$1(3)\times 10^{-13}$] \\
  $\tilde{P}'''$ [d$^{-2}$] & $-2.620(16)\times 10^{-16}$ &$[1(3)\times 10^{-16}]$  & [$-2(1)\times 10^{-16}$]  & [$4(4)\times 10^{-16}$]  \\
  $\tilde{P}''''$ [d$^{-3}$]&$2.67(15)\times 10^{-20}$&&&\\
  $\tilde{P}'''''$ [d$^{-4}$]&$[8(3)\times 10^{-24}$]&&&\\
  $\tilde{P}/\tilde{P}'$ [yr] &$5.1\times10^{5}$& $3.8\times10^{5}$&$1.1\times10^6$ & $1.1\times10^6$\\
  $n/n_{\rm phot}$&18617/17922&4394/3656&27656/27373&13591/13591 \\
  $\sigma$  [mag]&0.0056&0.0050&0.0036&0.018\\
  $g$ &276&73&56&57\\
  $n/\sigma^2/10^6$ &600&180&2100&42\\
   obs. int.& 1949-2016 & 1976-2016 & 1974-2017 & 1975-2014 \\
   $n_{\rm rev}$& 47\,225 &9556  & 12\,915 &6514 \\
   ph. ampl.&0.62&$>0.51$&$>0.26$&$>0.11$\\
   $\beta_{20}$ & 5.734$\times 10^{7} $ & $9.111\times10^6$ & $1.120\times10^6$& $3.256\times10^6$\\
   $\beta_{21}$ & -1.002$\times 10^{4} $ & $-4.486\times10^3$ & $-9.418\times10^3$& $-3.001\times10^3$\\
   $\beta_{30}$ & 6.765$\times 10^{11} $ & $8.060\times10^9$ &$-2.284\times10^9$ & $6.139\times10^9$\\
   $\beta_{31}$ & 1.5570$\times 10^{8} $ & $1.005\times10^7$ & $2.198\times10^7$ &$-6.706\times10^8$\\
   $\beta_{32}$ & -2.1819$\times 10^{4} $ & $-5.371\times10^3$ &$-7.378\times10^3$ & $-4.887\times10^3$\\
   $\beta_{40}$ & 3.025$\times 10^{15} $ &$-4.001\times10^{13}$& $-8.834\times10^{12}$ &$3.961\times10^{11}$\\
   $\beta_{41}$ & 3.5026$\times 10^{12} $ &$4.055\times10^{10}$ & $7.873\times10^{10}$& $1.3518\times10^{10}$\\
   $\beta_{42}$ & -1.979$\times 10^{7} $ &$1.035\times10^{7}$ &$6.940\times10^6$ & $-5.4553\times10^6$\\
   $\beta_{43}$ &$-3.4066\times 10^{4} $ &$-6.283\times10^3$ &$-6.4403\times10^3$  & $1.120\times10^6$\\
   $\beta_{50}$ & 7.999$\times 10^{20} $ &  & & \\
   $\beta_{51}$ & -3.1565$\times 10^{16} $ &  &  &\\
   $\beta_{52}$ & -3.1346$\times 10^{13} $ &  &  & \\
   $\beta_{53}$ & -9.4321$\times 10^{7} $ &  &  & \\
   $\beta_{54}$ & -3.9449$\times 10^{4} $ &  &  & \\
  \hline\hline
\end{tabular}
\end{table}
Unfortunately, we do not have reliable statistics on the percentage of rotationally unstable stars among various types of mCP stars. This complicates our further speculation on the nature of the observed period variations. Nevertheless, it would be inspiring to discuss possible explanations and expectations more carefully.

In the following, we shall discuss the particular results of the period analyses, many of which are presented in Table\,\ref{tab}. The main source of the phase information is the photometry which is most plentiful and relatively accurate.

The best-observed star, judging according to the ratio of the total number of observations $n$ to the square of the mean uncertainty $\sigma$, $n/\sigma^2$, is apparently \sig, which is mostly due to the unprecedentedly good and numerous observations from the MOST satellite \citep{townmost}. Very well observed is also \vir\ thanks to long-time photometric monitoring of the star by Diana Pyper and Saul Adelman \citep[][and many others]{pyper13}.

The only star showing verifiably sinusoidal long-term changes in its period is \vir\ which has been monitored during the whole cycle of about 66 years. For expressing a satisfactory polynomial model of the phase function (see Sect.\,\ref{ortogpolmodel}), a quintic parabola is needed, in the case of \vir, a cubic parabola for \ori\ and a simple square parabola for the other two stars. Higher terms seem to be nearly negligible. Unfortunately, we cannot dig up more information on possible cyclicity in all the stars, excluding \vir. It is of particular interest that the mean rate of the period is rising in all the stars, although the rotation of both \vir\ and \ori\ is currently accelerating.

Some physical information, excluding precession as the explanation of period changes in \vir, \ori, and \sig\ is the amplitude of phase change $A_{\varphi}$. In the case of \vir\ we can read the amplitude $A_{\varphi}$ directly from its \oc\ diagram -- 0.62 of the period (see Fig.\,\ref{duoCU}); for the other stars we can estimate at least its lower limit according to a simple relation: $A_{\varphi}>\tilde{P}'/2 n_{\rm {rev}}^2$, where $n_{\rm {rev}}$ is the number of revolutions monitored. The limit for \ori\ is 0.55 of the revolution, 0.26 in \sig\ and only 0.11 for \cir. This value is allowed by the model of free precession \citep[see appendix in][]{mik901}. However, the purely quadratic course of the modulation of the phase function indicates that the true phase amplitude is rather larger.

The nature of the period variations of the discussed stars is not known up to now, and we cannot exclude that their reasons may be different. The present analy-sis supports the idea \citep[more broadly investigated in][]{miknat} that the tiny, nearly linear period changes of \cir\ can be due to precession of the magnetically distorted star, while the nearly sinusoidal period oscillations of at least \vir\ might be interpreted as a consequence of internal waves disseminating within the magnetic rotating stars. Period changes in hot mCP stars with stellar wind escaping from their extended magnetosphere can be explained by angular momentum loss (\sig), modulated by the gradual reconfiguration of the magnetic field firmly connected with the surface layers (\ori).

\section{Conclusions}

We have presented a novel and commonly applicable method of period analysis based on the simultaneous exploitation of all available observational data containing phase information. This phenomenological method can monitor gradual changes in the observed instantaneous period very efficiently and reliably.

We have presented up to date results of the period monitoring of \ori, \vir, \sig, and \cir, known to be mCP stars changing their observed periods and have discussed the physics of this unusual behavior. To compare the period behavior of those stars, we treated their data with an orthogonal polynomial model, which we specifically develop for this purpose.

We have confirmed period variations in all investigated stars and shown that they reflect real changes in the angular velocity of outer layers of the stars fastened by their global magnetic fields.

However, the nature of the observed rotational instabilities has remained elusive up to now. The discussed group of mCP stars is inhomogeneous to such extent that each of the stars may experience a different cause for its period variations.

\acknowledgements
The research was supported by the project GA\,\v{C}R 16-01116S. The author thanks to Stefan H\"{u}mmerich and Miroslav Jagelka for their careful reading of the manuscript and improving its language.

The author of the review paper is also indebted to his close collaborators and all who provided him with data on investigated mCP stars, particularly to S.\,A. Adelman,  K.~Bernhardt, M.~Chrastina, S.~de Villiers, M.~Drozd, G.\,W.~Henry, S.~Hubrig, S.~H\"{u}mmerich, M. Jagelka, J.~Jan\'ik, O.~Kochukhov, J. Krti\v{c}ka, D.\,O. Kudryavtsev, P. Kurf\"{u}rst, R. Kuschnig, J.~Landstreet, J. Li\v{s}ka, T.~ L\"{u}ftinger, M.~Netopil, M.~Oksala, E.~Paunzen, T.~Pribulla, M.~Prv\'{a}k, D.~Pyper, T.~Rivinius,  A.~ Reiners,  I.\,I.~ Romanyuk, T.~Ryabchikova, E.~Semenko, M.~Shultz, D. Shulyak, N.~Sokolov, K.~Stepie\'n, G.~Sz\'{a}sz, R.H.D. Townsend, C.~ Trigilio, M.~Va\v{n}ko, G.\,A.~Wade, M.~Zejda, P.~Zielinski, J.~Zverko, P.~Zv\v{e}\v{r}ina, J. \ziga, and many others.


\begin{thebibliography}{}
\bibitem[Belopolsky(1913)]{belopol} Belopolsky, A.: 1913, {\it \an}, {\bf 195}, 159
\bibitem[Bevington \& Robinson(2003)]{bevington} Bevington, P. R., \& Robinson, D. K., in {\it Data Reduction and Error Analysis for the Physical Sciences} (3rd ed.), ISBN-10: 0072472278 , McGraw-Hill, 2003
\bibitem[Catalano \& Leone(1993)]{cale93} Catalano, F. A. \& Leone, F.: 1993, {\it \aaas}, {\bf 100}, 319
\bibitem[Cidale et al.(2007)]{cid} Cidale, L. S., Arias, M. L., Torres, A. F., Zorec, J., Fr\'emat, Y., \& Cruzado, A.: 2007, {\it \aaa} {\bf 468}, 263
\bibitem[Deutsch(1952)]{deutsch} Deutsch, A. J.: 1952, {\it \apj}, {\bf 536}
\bibitem[Donati \& Landstreet(2009)]{dolan} Donati, J.-F. \& Landstreet, J.D.: 2009, {\it \ar}, {\bf 47}, 333
\bibitem[Dubath et al.(2011)]{dubath} Dubath, P., Rimoldini, L., S\"{u}veges, M., Blomme, J., L\'{o}pez, M., Sarro, L. M., De\,Ridder, J., Cuypers, J., Guy, L.,  Lecoeur, I.  et al.: 2011, {\it \mnras}, {\bf 414} 2602
\bibitem[Farnsworth(1932)]{farn} Farnsworth, G.: 1932, {\it \apj}, {\bf 75}, 364
\bibitem[Hartkopf et al.(1989)]{hart} Hartkopf, W. I., McAlister. H. A., \& Franz, O. G.: 1989, {\it \aj}, {\bf 98}, 1014
\bibitem[Hesser et al.(1977)]{hesser} Hesser, J. E., Ugarte, P. P., \& Moreno, H.: 1977, {\it \apj}, {\bf 216}, L31
\bibitem[Hubrig et al.(2006)]{hub06} Hubrig, S., North, P., Sch\"{o}ller, M., \& Mathys, G.: 2006, {it\ \an}, {\bf 327}, 289
\bibitem[Hunger et al.(1989)]{hunger} Hunger, K., Heber U., \& Groote, D.: 1989, {\it \aaa}, {\bf 224}, 57
\bibitem[Jagelka \& \mik(2015)]{jagmik} Jagelka, M. \& \mik, Z.: 2015, in {\it Physics and Evolution of Magnetic and Related Stars}, Eds.: Y. Y. Balega, I. I. Romanyuk, D. O. Kudryavtsev, {\it ASP Conf. Ser.}, {\bf 494}, 230
\bibitem[Jan\'{i}k et al.(2011)]{jan}  Jan\'{i}k, J., \mik, Z., Sz\'{a}sz, G., Zejda, M., Zv\v{e}\v{r}ina, P., Zverko, J. \& \ziga, J.: 2011, in {\it Magnetic Stars},  Proceedings of the International Conference, SAO RAS 2010, Eds: I. I. Romanyuk and D. O. Kudryavtsev, 476
\bibitem[Khokhlova et al.(2000)]{choch} Khokhlova V. L., Vasilchenko D. V., Stepanov V. V., \& Romanyuk, I. I.: 2000, {\it Astron. Letters}, {\bf 26}, 177
\bibitem[Kochukhov \& Bagnulo(2006)]{kobacu} Kochukhov, O. \& Bagnulo, S.: 2006, {\it \apj}, {\bf 726}, 24
\bibitem[Kochukhov et al.(2011)]{ko901} Kochukhov, O., Lundin, A.,
    Romanyuk, I., \& Kudryavtsev, D. O.: 2011, {\it \aaa}, {\bf 450}, 763
\bibitem[Krti\v{c}ka(2014)]{krtwind} Krti\v{c}ka, J.: 2014, {\it \aaa}, {\bf 564}, 70
\bibitem[Krti\v{c}ka et al.(2017)]{krtosc} Krti\v{c}ka, J., \mik, Z., Henry, G. W., Kurf\"{u}rst, P., Karlick\'y, M.: 2017, {\it \mnras}, {\bf 464}, 933
\bibitem[Krti\v{c}ka et al.(2015)]{krtheta} Krti\v{c}ka, J., \mik, Z., L\"{u}ftinger, T., \& Jagelka, M.: 2015, {\it \aaa} {\bf 576}, 82
\bibitem[Krti\v cka et al.(2012)]{krtcu} Krti\v{c}ka, J., \mik, Z.; L\"{u}ftinger, T., Shulyak, D., Zverko, J., \ziga, J., \& Sokolov, N. A.: 2012, {\it \aaa}, {\bf 534}, 5
\bibitem[Krti\v{c}ka et al.(2007)]{krt901} Krti\v{c}ka, J., Mikul\'a\v{s}ek, Z., Zverko, J., \& \ziga, J.: 2007, {\it \aaa}, {\bf 470}, 1089
\bibitem[Kuschnig et al.(1999)]{kusch} Kuschnig, R., Ryabchikova, T. A., Piskunov, N. E., Weiss, W. W., \& Gelbmann, M. J.: 1999, {\it \aaa}, {348}, 924
\bibitem[Li\v{s}ka et al.(2016)]{liska} Li\v{s}ka, J., Skarka, M., Zejda, M., \mik, Z., \& de Villiers, S. N.: 2016, {\it \mnras}, {\bf 459}, 4360
\bibitem[L\"{u}ftinger et al.(2010)]{luft} L\"{u}ftinger, T., Kochukhov, O., Ryabchikova, T., et al.: 2010, {\it \aaa}, {\bf 509}, A71
\bibitem[Manfroid \& Renson(1980)]{mare80} Manfroid, J. \& Renson, P.: 1980, {\it \ibvs}, 1824
\bibitem[Manfroid \& Renson(1983)]{mare83} Manfroid, J. \& Renson, P.: 1983, {\it \aaas}, {\bf 51}, 267
\bibitem[Mathys \& Manfroid(1985)]{mama85} Mathys, G. \& Manfroid, J. 1985, {\it \aaas}, {\bf 60}, 17
\bibitem[Mathys(1988)]{mathys} Mathys, G.: 1988, {\it \aaa}, {\bf 189}, 179
\bibitem[\mik(2007b)]{mikort} \mik, Z.: 2007b, {\it Odessa Astron. Publ.}, {\bf 20}, 138
\bibitem[\mik(2007c)]{apca} \mik, Z.: 2007c, {\it Astron. Astrophys. Trans.}, {\bf 26}, 63
\bibitem[Mikul\'a\v{s}ek(2015)]{mikecl} Mikul\'a\v{s}ek, Z.: 2015,
  {\it \aaa}, {\bf 584}, A8
\bibitem[\miket(2007c)]{mikan} \mik, Z., Jan\'ik, J., Zverko, J., \ziga, J., Zejda, M., Netolick\'y, M., \& Va\v{n}ko, M.: 2007c, {\it \an}, {\bf 328}, 10
\bibitem[\miket(2015a)]{mik15} Mikul\'a\v{s}ek, Z., Jan\'ik, J., Krti\v{c}ka, J., Zejda, M., \& Jagelka, M.: 2015a, in {\it Physics and Evolution of Magnetic and Related Stars}, Eds.: Y. Y. Balega, I. I. Romanyuk, D. O. Kudryavtsev, {\it ASP Conf. Ser.}, {\bf 494}, 189
\bibitem[\miket(2017)]{miknat} \mik, Z., Krti\v{c}ka, J., Jan\'{i}k, J., Henry, G. W., Zejda, M., Shultz, M., Paunzen, E., \& Jagelka, J.: 2017, in {\it Stars: From collapse to collapse}, Eds: D. Kudryavtsev, I. I. Romanyuk, I. Balega, in press
\bibitem[\miket(2007a)]{mik07} Mikul\'a\v{s}ek, Z., Krti\v{c}ka, J., Zverko, J. et al.: 2007, in Active OB-Stars: Laboratories for Stellar and Circumstellar Physics, ASP Conference Series, Vol. 361, Proceedings of the conference, Sapporo, Japan. Eds. S. Stefl, S. P. Owocki, and A. T. Okazaki. {\it Astronomical Society of the Pacific}, 2007, 466
\bibitem[Mikul\'a\v{s}ek et al.(2008)]{mik901} Mikul\'a\v{s}ek, Z., Krti\v{c}ka, J., Henry, G. W., Zverko, J., \ziga, J. et al.: 2008, {\it \aaa}, {\bf 485}, 585
\bibitem[Mikul\'a\v{s}ek et al.(2011a)]{mik11} Mikul\'a\v{s}ek, Z., Krti\v{c}ka, J., \& Henry, G. W. et al.: 2011a, {\it \aaa}, {\bf 534}, L5
\bibitem[Mikul\'a\v{s}ek et al.(2011b)]{unstead} Mikul\'a\v{s}ek, Z., Krti\v cka, J., Jan\'ik J., Zverko, J., \ziga, J., Zv\v{e}\v{r}ina, P., \& Zejda, M.: 2011b, in {\it Magnetic Stars},  Proceedings of the International Conference, SAO RAS 2010, Eds: I. I. Romanyuk and D. O. Kudryavtsev, 52
\bibitem[Mikul\'a\v{s}ek et al.(2014)]{mikmos} Mikul\'a\v{s}ek, Z., Krti\v{c}ka, J., Jan\'ik, J., Zejda, M., Henry, G. W., Paunzen, E., \ziga, J., \& Zverko, J.: 2014, in {\it Putting A Stars into Context: Evolution, Environment, and Related Stars}, eds. G. Mathys, E. Griffin, O. Kochukhov, R. Monier, G. Wahlgren, Pero, Moscow, 270
\bibitem[\miket(2015b)]{mikper} Mikul\'a\v{s}ek, Z., Paunzen, E., Netopil, M., \& Zejda, M.: 2015b, in {\it Physics and Evolution of Magnetic and Related Stars}, Eds.: Y. Y. Balega, I. I. Romanyuk, D. O. Kudryavtsev, {\it ASP Conf. Ser.}, {\bf 494}, 320
\bibitem[\miket(2016)]{mikic} \mik, Z., Paunzen, E., Zejda, M., Semenko, E., Bernhard, K., H\"{u}mmerich, S., Zhang, J., Hubrig, S., Kuschnig, R., Jan\'{i}k, Jan, \& Jagelka, M.: 2016, {\it Bulg. Astron. J.} {\bf 25} 19
\bibitem[\miket(2009)]{mikper} \mik, Z., Szasz, G., Krti\v{c}ka, J., Zverko, J., \ziga, J., Zejda, M., \& Graf, T.: 2009, ArXiv:0905.2565
\bibitem[Mikul\'a\v{s}ek \& Zejda(2013)]{mikzej} Mikul\'a\v{s}ek, Z. \&
    Zejda, M., in {\it\'Uvod do studia prom\v{e}nn\'{y}ch hv\v{e}zd},
    ISBN 978-80-210-6241-2,  Masaryk University, Brno 2013
\bibitem[\miket(2004)]{mik04} \mik, Z., Zverko, J., \ziga, J., \& Jan\'ik, J.: 2004, in {\it The A-Star Puzzle}, Eds. J. Zverko, J. \v{Z}i\v{z}\v{n}ovsk\'y, S. J. Adelman, \& W.W. Weiss, IAU Symposium, {\bf 224}, 657
\bibitem[\miket(2007b)]{mik07b}Mikul\'a\v{s}ek, Z., Zverko, J., Krticka, J.,  \& Zejda, M.:2007b, in Magnetic stars, eds. Iosif Romanyuk \& Dmitry Kudryavtsev,
\bibitem[Oksala et al.(2015a)]{oxaiaus} Oksala, M. E., Kochukhov, O., Krti\v{c}ka, J., Prv\'{a}k, M., \& \mik, Z.: 2015  2015IAUS..307..348O
\bibitem[Oksala et al.(2015b)]{oxa15} Oksala, M. E., Kochukhov, O., Krti\v{c}ka, J., Townsend, R. H. D., Wade, G. A., Prv\'{a}k, M., \mik, Z., Silvester, J., \& Owocki, S. P.: 2015b, {\it \mnras}, {\bf 451}, 2015
\bibitem[Oksala \& Townsend(2007)]{oxa} Oksala, M. \&  Townsend, R. H. D.: 2007, in Okazaki A. T., Owocki S. P., Stefl S., eds, ASP. Conf. Ser. 361: Active OB-Stars: Laboratories for Stellar and Circumstellar Physics, 476
\bibitem[Pedersen(1979)]{ped} Pedersen, H.: 1979, {\it \aaas}, {\bf 35}, 313
\bibitem[Press et al.(2002)]{press} Press, W. H., Teukolsky, S. A., Vetterling, W. T. \& Flannery, B. P.: 2002,  {\it Numerical recipes in C++ : the art of scientific computing}, by William H. Press, ISBN: 0521750334
\bibitem[Pyper \& Adelman(2004)]{pyper04} Pyper, D. ,M., Adelman, S.\,J.: 2004, in {\it The A-Star Puzzle, IAU Symposium No. 224}, eds. J. Zverko, J. \v{Z}i\v{z}\v{n}ovsk\'{y}, S.\,J. Adelman, \& W.\,W. Weiss (Cambridge University Press, Cambridge){}, 307
\bibitem[Pyper et al.(1997)]{pyper97} Pyper, D. ,M., Ryabchikova, T., \&
    Malanushenko, V.: 1997, {\it \baas}, {\bf 29}, 811
\bibitem[Pyper et al.(1998)]{pyper98} Pyper, D. M., Ryabchikova, T.,  \& Malanushenko, V. et al.: 1998, {\it Astron. Astrophys.}, {\bf 339}, 822
\bibitem[Pyper et al.(2013)]{pyper13} Pyper, D. M., Stevens, I. R., Adelman, S. J.: 2013, {\it \mnras}, {\bf 431}, 2106
\bibitem[Ravi et al.(2010)]{ravi} Ravi, V., Hobbs, G., Wickramasinghe,
    D., Champion, D. J., \& Keith M.: 2010, {\it \mnras}, {\bf 408}, 99
\bibitem[Rice et al.(1989)]{rice} Rice, J. B., Wehlau, W.H., \& Khokhlova V. L.: 1989, {\it \aaa}, {\bf 208}, 179
\bibitem[Reiners et al.(2000)]{reiners} Reiners, A., Stahl, O., Wolf, B., Kaufer, A., \& Rivinius, T.: 2000, {it \aaa}, {\bf 363}, 585
\bibitem[Shore \& Adelman(1976)]{shoade} Shore, S. N. \& Adelman, S. J.:
    1976, {\it \apj}, {\bf 209}, 816
\bibitem[Shulyak et al.(2010)]{shul} Shulyak, D., Krti\v{c}ka, J., \mik, Z., Kochukhov, O., \& L\"{u}ftinger, T.: 2010,{\it \aaa}, {\bf 524}, A66
\bibitem[Silvester et al.(2014)]{silvester} Silvester, J., Kochukhov, O., \& Wade, G. A.: 2014, {\it \mnras}, {\bf 440}, 182
\bibitem[Silvester et al.(2015)]{silva15} Silvester, J., Kochukhov, O., \& Wade, G. A.: 2015, {\it \mnras}, {\bf 453}, 2163
\bibitem[St\c{e}pie\'n(1998)]{step} St\c{e}pie\'n, K.: 1998, {\it \aaa}, {\bf 337}, 754
\bibitem[Thompson \& Landstreet(1985)]{thomla} Thompson~I.\,B. \&
    Landstreet~J.\,D.: 1985, {\it \apj}, {\bf 289}, 9
\bibitem[Townsend et al.(2010)]{town} Townsend, R. H. D., Oksala, M. E.,
    Cohen, D. H., Owocki, S. P., \& ud--Doula, A.: 2010, {\it \apj}, {\bf 714}, 318
\bibitem[Townsend et al.(2013)]{townmost} Townsend, R. H. D., Rivinius, Th., Rowe, J. F., Moffat, A. F. J., Matthews, J. M.; Bohlender, D., Neiner, C., Telting, J. H., Guenther, D. B., Kallinger, T. et al.: 2013, {\it \apj}, {\bf 769}, 33
\bibitem[Trigilio et al.(2000)]{trigi00} Trigilio, C., Leto, P., Leone, F., Umana, G., \&  Buemi, C.: 2000, {\it \aaa}, {\bf 362}, 281
\bibitem[Trigilio et al.(2008)]{trigi} Trigilio, C., Leto, P., Umana,
    G., Buemi, C. S., \& Leone, F.: 2008, {\it \mnras}, {\bf 384}, 1437
\bibitem[Trigilio et al.(2011)]{trigi11} Trigilio, C., Leto, P., Umana,
    G., Buemi, C. S., \& Leone, F.: 2011, {\it \apj}, {\bf 739}, L10
\bibitem[ud-Doula et al.(2009)]{ud} ud-Doula, A., Owocki, S. P., \& Townsend, R. H. D.: 2009, {\it \mnras}, {\bf 392}, 1022
\bibitem[Vogt \& Faundez(1979)]{vofa79} Vogt, N. \& Faundez, A. M.: 1979 {\it\aaa}, {\bf 36}, 477
\bibitem[Wade et al.(2000)]{wade00} Wade, G.A., Donati, J.-F., \& Landstreet, J.D.: 2000, {\it \mnras}, {\bf 313}, 851
\bibitem[\ziga\ et al.(2000)]{zigasx} \ziga, J., Schwartz, P., \& Zverko, J.: 2000, {\it \ibvs} {\bf 4835}

\end{thebibliography}
\end{document}